\documentclass[11pt, oneside]{article}   	% use "amsart" instead of "article" for AMSLaTeX format
\usepackage{geometry}                		% See geometry.pdf to learn the layout options. There are lots.
\usepackage{amsmath}
\usepackage{amssymb}
\usepackage[utf8]{inputenc}
\usepackage[english]{babel}
\usepackage{graphicx}

\usepackage{ dsfont }
\usepackage{natbib}

\usepackage{caption}
\usepackage{subcaption}

\usepackage[parfill]{parskip}    		% Activate to begin paragraphs with an empty line rather than an indent
\usepackage{graphicx}				% Use pdf, png, jpg, or eps§ with pdflatex; use eps in DVI mode
\usepackage{amssymb}

\usepackage{authblk}
\title{Status hierarchy and group cooperation:  A generalized model}

\author[1]{Hsuan-Wei Lee}
\author[1,2]{Yen-Ping Chang}
\author[1]{Yen-Sheng Chiang\thanks{Corresponding author. Email: chiangys@gate.sinica.edu.tw}}
\affil[1]{Institute of Sociology, Academia Sinica, Taiwan}
\affil[2]{Department of Educational Psychology and Counseling, National Tsing Hua University, Taiwan}
\date{}		         	% Activate to display a given date or no date

\begin{document}
\maketitle
\begin{abstract}
In a refreshing mathematical investigation, Mark (2018) shows that status hierarchy may facilitate the emergence of cooperation in groups. Despite the contribution, the present paper notes that there are limitations in Mark’s model that makes it less realistic than it could in explaining real-world experiences. Consequently, we present a more generalized modified framework in which his model is a special case, by developing and introducing a new hierarchy measure into the model to estimate the cooperation level in a set of hierarchical structures omitted in Mark's work yet common in everyday life--those with multiple leaders. We derived the conditions under which cooperation can emerge in these groups, and verified our analytical predictions in agent-based computer simulations. In so doing, not only does our model elaborate on its predecessor and support Mark's general prediction. For theory, our work further reveals two novel phenomena of group cooperation: Both the relative number of cooperators to defectors in groups and the assortativity among these different roles can backfire; they are not always the higher, the better for cooperation to thrive. For methodology, the hierarchy measure developed and our model using the measure may also be applied in future research on a wide range of related topics.
\end{abstract}

\section{INTRODUCTION}

Group cooperation and status inequality have each received longstanding attention in sociology. While they are often discussed in independent lines of research \citep{kollock1998social,correll2006expectation, fehr2007human, sauder2012status}, scholars have recently started investigating both topics in an integrated view \citep{simpson2015beyond}. Specifically, researchers argue that hierarchy may benefit the emergence of cooperation in groups over evolutionary history against defection \citep{hechter1987principles, whitmeyer2007prestige, willer2009groups, mark2018status}, thus linking two lines of investigation into one. Take the leader-follower status for example. Ample evidence from field and experimental work have shown that people behave more cooperatively in groups with leaders than without \citep{luo2007elections, olken2010direct, hamman2011experimental, grossman2012impact, bolsen2014voters}. 

Compared to the fruitful empirical demonstrations of the phenomenon, theoretical explanations for the phenomenon--that is, the mechanisms through which status hierarchy induces group cooperation in evolution--have nonetheless remained relatively untouched. Tackling this gap in the literature, we believe the evolutionary game (EVG) theory may contribute to the advance of theory. EVG is a tool that not only biologists \citep{nowak2006five} but also social scientists \citep{bendor2001evolution, bergstrom2002evolution} utilize to formalize explanations of the emergence of human behavior and social institution, such as cooperation and status hierarchy. In particular, EVG aims to identify the conditions under which the behavior of interest--represented by a strategy in a game--comes into being and remains stable against alternative behaviors in a given social context--a set of game settings. The emergence of the target behavior is measured by its evolutionary fitness, in the sense that the more benefits accrued to the behavior, the more (likely) it would be adopted by an actor, or by more actors. As far as cooperation is concerned, scientists have used the EVG framework to pin down an array of mechanisms by which the free-riding problem can be mitigated in cooperation dilemmas \citep{nowak2006five}. We hence follow the literature and apply EVG to examine whether and how status hierarchy facilitates cooperation in groups.

Guiding our investigation, the paper of \citep{mark2018status} provides a pioneering and refreshing answer to our research question. In his model, conditional cooperators--called status cooperators--first engage in a pursuit of leadership. They then cooperate, fully, if one and only one of them surfaces to take the leader position; they defect otherwise if no leader is chosen. In comparison to cooperators, status defectors never pursue the leader role, nor do they cooperate with others. Using this EVG, Mark demonstrates that, even though defectors seem to be the advantaged as they always free-ride on the efforts of cooperators, it is possible for status cooperators to outperform defectors in evolution history as long as one of cooperators becomes the leader and make their fellow cooperators to cooperate. Finally, \citep{mark2018status} reports that this status-dependent cooperation behavior is even more likely to emerge against defection when the group size is smaller than larger, and when the group formation is more assortative (i.e., actors are disproportionately likely to interact with same-role others) than random.

Building off its merits and insights, we however note that Mark’s EVG is limited too and, specifically, does not take into account the situation in which multiple leaders emerge from cooperators and lead the group in a parallel fashion. As the late sociologist Roger Gould argued, ``It is evident that most social hierarchies lie between the two extremes of complete equality [i.e., no leader selected] and winner-take-all inequality [i.e., one leader only]." \citep[p.~1149]{gould2002origins}. In short, real-world status difference is not always, following Mark's terms, clear and apparent. We therefore tackle the issue in the present paper, therefore extending the investigation of Mark.

In particular, below we show that stipulating status structure in groups to be either no-leader or single-leader makes Mark's model sometimes underestimate the possibility that cooperation can emerge in groups, and other times overestimate the same possibility. Practically, we build upon his model by introducing into the model a continuous measure of status ``hierarchicalness" to capture the influences of multi-leader status structures excluded in Mark's work. The measure allows us to extrapolate the extent to which group members cooperate at any given level of hierarchicalness. In other words, we present a more generalized framework in which Mark’s original model is a special case. In comparison to Mark's (2018) reliance on analytical solutions, we also verify our EVG by computer simulations; the code is provided online.

In addition, two novel phenomena not revealed in Mark's work are reported using our more generalized model. For one, according to Mark's model equations, his model would predict that, in hierarchy, the more status cooperators in a group relative to defectors, the more viable cooperation is a behavioral strategy in evolution against defection. This extension of Mark's work is appealing, and we verified it in the present paper. Nonetheless, we further show that the prediction is true only when the existence of hierarchy is an either-or issue as in Mark's model, that is, when there is either a clear hierarchy of only one leader, or no hierarchy and no leader at all. By contrast, when hierarchicalness is conceptualized as a more nuanced continuous construct as in our generalized framework, the cooperator-to-defector ratio only helps with cooperation up to a point, and then it backfires. In other words, we report the possibility of an ironic effect of having too many cooperators in a group on generating cooperation in the group. 

Secondly, we investigate the interdependence between the relative number of cooperators to defectors and the assortativity between the two roles. To the issue, Mark demonstrates that, much like the relative number of cooperators, assortativity increases the evolutionary strength of cooperation against defection. We replicated the prediction in our generalized model. However, we also found that the benefit of assortativity comes with a cost: It may compromise the benefit of the number of cooperators. Sometimes, when the degree of assortativity is large enough, it may even reverse the effect of the cooperator-to-defector ratio, making it detrimental but not beneficial to the evolution of cooperation. The implication, then, is that not only can increase the ratio of cooperators in a group backfires at some point, but the increase might also be simply destructive at any point given high assortativity, despite that assortativity is commonly thought of as a ``protector" of cooperation in the literature. Below, we detail our generalized group cooperation model with a rigorous formalism and nuanced cooperative properties.

\section{SOCIAL HIERARCHY AND COOPERATION: AN EVOLUTIONARY ACCOUNT}

\subsection{Status hierarchy}

Status inequality is not only ubiquitous in human societies, but also prevalent in the societies of many species of social animals \citep{chase1985sequential, sapolsky2005influence, sauder2012status}. Biologists argue that social hierarchy, particularly the dominance type, is a result of constant competition for valuable resources such as food and mating opportunities \citep{hobson2019differences}. For humans, social hierarchy and status inequality are further suggested as a cause of social and health problems \citep{hemingway1997impact, burns2014income}. Given the negative consequences of inequality, it is thus puzzling why it persists in the real world.

At least two reasons have been proposed to address this puzzle. First, scholars argue that, even though humans are mostly fairness-minded, perfect equality is never their ultimate goal; people do desire some if not overly steep hierarchy \citep{starmans2017people}. Supporting this idea, evidence suggests that people would accept a certain degree of inequality in exchange for other types of personal benefits \citep{kuziemko2014last}. Research has also shown that once a status hierarchy is formed, people not only tend to live with it but also, sometimes, strive to protect this status quo \citep{xie2017rank}. Secondly, scholars propose that social hierarchy can do more than the lesser evil and actually be beneficial to social coordination and cooperation. For instance, evolutionary psychologists attribute the emergence of leadership and, therefore, the leader-follower status disparity among animals and humans to its potential function of exerting social influences, especially in but not limited to risky situations and critical group-wide challenges such as hunting for food and fighting with rivalry tribes \citep{van2006evolutionary}. In support of the argument, experimental studies have shown that people tend to behave more cooperatively with leaders than without \citep{luo2007elections, olken2010direct, hamman2011experimental, grossman2012impact, bolsen2014voters}. It is also found that cooperative actors receive more status approval and are elected as leaders more often than do non-cooperative members of the group \citep{willer2009groups}. Because of these collective functions, a social hierarchy may therefore hold in human societies.  

\subsection{Mark’s (2018) model: contributions, limitations, and our extensions} 

In his model, Mark (2018) examines one specific hierarchical structure in how likely cooperation would emerge in it over an evolutionary game. The structure features two types of actors: status cooperators ($SC$) and defectors ($D$). $SC$s pursue the leader position whereas the $D$s do not, and the leadership is determined in a process of intent signaling: Every SC signals whether they are interested in being the leader with a given probability, and the hierarchy is settled when exactly one SC shows the interest and becomes the leader. The process repeats if multiple $SC$s express the interest until one and the only leader is chosen or none of $SC$s signals the intent to lead.

Despite its originality, Mark’s model is however limited and overly simplified, because single-leader hierarchy is merely one of the various hierarchical structures we witness in the real world. As such, the goal of the present paper is to generalize Mark's work over an extra dimension of the leader-follower structure: the number of leaders. In Mark's model, the dimension takes only two extreme values: 0 for no hierarchy at all, and 1 for total dominance of the one leader. By contrast, we first developed and incorporated into our model a measure of parallel competing leadership allocated to multiple leaders, and then derived the implications of this attenuated status difference for cooperation in groups. Below, let us first quickly summarize the part of Mark's model that we built ours on. 

In Mark's original model, there are four events happening in one iteration of the game: (1) the determination of the group; (2) preplay communication for leader selection; (3) cooperators' decisions of whether to contribute to the group; (4) the determination of next generation of status cooperators and defectors based on the payoffs of the current generation. Mark states that, in the phase of preplay signaling and formation of a status hierarchy, in each round of signaling, every status cooperator signals the/high intend to lead the group with probability $\frac{1}{n}$ where $n$ is the group size, and no/low such intend with the complementary probability $1-\frac{1}{n}$; every defector signals low intend. The signaling repeats for rounds until a clear status hierarchy is reached, i.e., when there is no player at the top, leader level, or when there is exactly one player at the level. Once the status hierarchy is obvious with 0 or 1 player at the top level, the cooperators continue to decide whether to contribute to the group. In Mark's own words, ``Each of the status cooperators in a group signals one or more times until either a status hierarchy is apparent (i.e., exactly one member of the group signaling high and all others signaling low) or a collective lack of confidence and leadership is indicated (i.e., each member of the group signaling low on the first round of signaling)." Following this setting, it is then reasonable to assume that when there are more than 1 player at the top level, the group continues on preplay signaling, and the process concludes when 0 or 1 player being at the top level.

To incorporate multi-leader structures into Mark's model, we simply break his requirement for hierarchy and allow hierarchies of any number of leaders nominated in the initial round of preplay communication to proceed to the next phase of the cooperators' deciding whether to contribute to the group. Supporting the design empirically, organizational hierarchy is known to take diverse forms in the real world, well beyond the either-or, one-or-no-leader type studied by Mark. In professional sports, for instance, it is common to see multiple star players on the same team and, indeed, on the court at the same time. In the business domain, it is also normal to have multiple managers in the executive committee of a company, each of whom is ranked equally in the status hierarchy. Overall, then, it seems crucial for us to include the continuous spectrum of structural hierarchicalness in between the two extreme ends of the signal- and the no-leader case. Our decision to not abandon multi-leader cases in the leader selection phase of the model also allows the model to analyze some issues that cannot be analyzed in Mark's work--e,g, the cost of repeated structure formation and leadership negotiation, which is implicitly assumed to be zero when multi-leader hierarchies are dropped without a cost in Mark's model yet is common and influential in the real world \citep{friedrich2009framework,dust2012and}.

Practically, if the signaling process generates multiple leaders in a round in our model, rather than dropping the round and starting all over as Mark's model would, we estimate the hierarchicalness of this--following his terminology--unapparent structure. We then estimate the downstream effects of this hierarchicalness in between that of no hierarchy at all and of total dominance of the one leader, on the upcoming cooperation decisions of cooperators in the group. To follow closely with Mark's model and preserve it as a special case in ours, we keep his model's feature that perfect equality generates no cooperation (from cooperators; defectors always defect) and total dominance generates unhesitating cooperation. Consequently, the issue that follows is how to first gauge the hierarchicalness of multi-leader structures when placing them in between the two extreme scenarios and then transform the hierarchicalness measure into cooperators' (un)certainty--the hesitation--to (not) cooperate.

To address the question, we visited the literature of measuring hierarchicalness of organizations, in search of an index that would meet the current needs. Specifically, a classic approach to measuring hierarchicalness ranks individuals by their dominance, following the idea that those who have a better winning rate and a larger winning differential over others in contests--or simply, stronger dominance--are expected to be ranked higher on the ladder. Traced back to the Bradley-Terry model \citep{bradley1952rank}, this family of methods has been modified to measure the status hierarchy in various organizations such as sports competitions \citep{massey1997statistical, colley2002colley} and peer relations between teenagers \citep{martin1998structures, levi2009formation}. Much like the first approach, an alternative way to measure social hierarchy is also built upon the same idea of dominance while taking a root in the social-network literature. From the network perspective, the dominance relation between a pair of actors can be represented by a directed edge stemming from actor $i$ to $j$, indicating that $i$ dominates $j$ in a diagram along a dimension of interest such as aggression, attraction, or prestige \citep{cheng2013two}. If the two actors are equal in status, their relationship can be represented by an undirected edge otherwise. As such, this network diagram approach allows researchers to use social network toolkits to measure the hierarchicalness of a group. For instance, Krackhardt proposed to measure the hierarchicalness of a group by the number of directed edges out of the total number of edges, directed or undirected, in the network \citep{krackhardt1994graph}. In ethology, researchers have also used the prevalence of cyclic subgraphs, wherein actor $i$ dominates $j$, who then dominates $k$, who in turn dominates $i$ \citep{shizuka2012social}, to gauge hierarchicalness. That is, if this kind of cycle is prevalent in a group, then the group is said to be less hierarchical than are the groups wherein $k$ does not dominate $i$ and is always fixed at the bottom. Finally, Mones et al. developed a hierarchicalness index by tracing the reachability of actors in the diagram of dominance relations \citep{mones2012hierarchy}. When a group is highly hierarchical, such as having a tree-like structure, the top actor--the root node--can reach all other nodes-actors (directly or indirectly) in the network, whereas the bottom node--the branches--can not reach any others, and those in between the top and the bottom can only reach those who are below themselves. In contrast, in the least hierarchical structure--a cycle--everyone can reach everyone else in time. The distribution of the reachability of nodes can consequently reveal the hierarchicalness of the group of the nodes.

Drawing from the literature, in this paper, we adopted the index of Mones et al. (2012) and modified it for Mark’s EVG (2018). We followed the tenet of Mark's work that ``the status hierarchy is the key to how status behavior promotes social order” (line 4-5, p. 1607), and then expanded on this idea by the hierarchicalness measure, so that the single-leader hierarchy in Mark's model becomes a special case of a more general model. Specifically, Mark reasons that ``when status behaviors combine to create a [clear, single-leader] status hierarchy, they help the members of a group [fully] overcome a social dilemma. When status behaviors do not [produce any leader and, thus, do not] create a status hierarchy, they do not promote cooperation [at all]” (line 5-8, p. 1607). Accordingly, we applied the chosen hierarchicalness measure to answer: If the single- and the no-leader structure represent the most and the least hierarchical structure respectively and they render total and zero cooperation among status cooperators respectively, what would be the level of cooperation in a group whose hierarchicalness is in between these two types of extremes and is measured by the index of \citep{mones2012hierarchy}? We detail our calculation below.

\section{THE MODEL}

The section is organized as follows. First, we introduce the hierarchicalness measure of \citep{mones2012hierarchy}. We then show how it is modified to fit into \citep{mark2018status} EVG to describe a more general relationship between status hierarchy and group cooperation. Along the way, we highlight the differences between Mark’s and our work, in the conditions under which cooperation could emerge. Lastly, for ease of reading, we follow the presentation of \citep{mark2018status} to first assume random mixing (Section 3.2) before introducing assortative mixing (Section 3.3) into the model.

\subsection{Hierarchical score and the cooperation probability $H_{n}(x)$}

As mentioned, there are four phases in one iteration in Mark's EVG: (1) the determination of the group; (2) preplay communication among group members; (3) cooperators' decisions to contribute to the group or not; (4) the determination of the next generation of status cooperators and defectors based on their payoffs from the current generation. We generalize this model by allowing multi-leader hierarchies to pass phase (2) and, therefore, forcing cooperators to make decisions on whether to contribute in such ``unapparent" structures in phase (3). Phases (1) and (4) are unchanged and, consequently, not discussed in detail below; we focus on phases (2) and (3).

As in Mark's model, there are two levels of hierarchy in our model to be formed from preplay communication: the leaders at the top and the followers at the bottom. In the process, every status cooperator similarly signals willingness to be a leader with probability $\frac{1}{n}$ and, hence, a lack of such intention with probability $1-\frac{1}{n}$, or $\frac{n-1}{n}$. If there is no player expressing the interest in taking the leadership role at the top level, following Mark, it is assumed that all status cooperators will not contribute to the group in the next phase of the game, as will the defectors at the bottom level. From here, however, Mark posits that status cooperators will only contribute and do so without hesitation if there is one and only leader emerging. If there are more, he requires the hierarchy formation to start over until a clear structure shows. To break from this requirement, we loosen the model design by incorporating the possibility that leadership is often shared by many in the real world \citep{selznick2011leadership, case2014divide, eslen2015toward}. Practically, if there are multiple players signaling high versus low intention to lead the group, instead of abandoning the round of leader selection, we retain the surfacing multi-leader structure and allow status cooperators to still contribute, though this time, with a likelihood--some hesitation. By so designing our EVG, then, not only is our model capable of studying more forms of hierarchy as we set out to achieve; there will be no issue of the hidden costs of failure for group members to agree upon a clear hierarchy (as is the case in Mark's work), because all multi-leader hierarchical statures are now considered a successful hierarchy formation and the costs of leader selection become a natural part of the game. Finally, it might be worth noting that the two extreme situations which Mark examines--that is, in which either one or zero leader is established--are taken into account in our model as two special cases out of all permitted hierarchical structures. Indeed, because Mark requires that all status cooperators cooperate when there is one leader and none of them cooperate when there is no leader, he sets up the boundaries of the cooperation probability of our model. That is, when the group is the most hierarchical, with one above all, all will follow and contribute with the highest cooperation probability of 1. By contrast, when the group is the least hierarchical, with all on the same flat level, none will contribute, or, will do so with the lowest cooperation probability of 0. Together, the two scenarios become the upper and the lower limit of our model, implying that all other less apparent hierarchical structures that we sought to include should sit between these two extremes of cooperation probability equal 0 or 1.

Following Mark's theorizing, we assume that, when there are multiple leaders at the top level and make the group's hierarchy obscure, the leaders' authority will become attenuated and, therefore, relatively ineffective in terms of motivating cooperation in the group. Importantly, not only is this assumption stems directly from Mark's reasoning; it has been supported in empirical studies. For instance, it has been documented that members of multi-leader teams simply need more resources, e.g., time, to communicate and to agree upon share goals \citep{rice2006opportunities}. This then increases the risk of failing to achieve consensus for which group members, especially subordinates, can work on together and cooperate with one another. Further, even if leaders can enclose distinct domains of leadership by, say, different specialties, allowing them to pursue independent objectives at the same time--much as what we modeled--runs on the possibility that they would get into each other's way, competing for common resources such as material and human resources \citep{friedrich2009framework}. On the other hand, though organizing leaders into a larger system may help them coordinate and therefore mitigate the problem \citep{dust2016multi}, it implicitly violates the idea of multi-leadership that we sought to investigate, as now the leaders are also followers under something or someone greater. In terms of mathematics, this means a reduction of the dimensionality of group leadership hidden behind a superficial constant number of leaders in the group. Consequently, we use the clarity of hierarchicalness of a group as the proxy to the likelihood that status cooperators can coordinate and then collaborate: The more clearly hierarchical a group, the more its cooperative members will cooperate. 

We adopt the measure of hierarchicalness developed by \citep{mones2012hierarchy}, because it bears the mathematical properties we were looking for. However, since this measure is defined for any arbitrary complete network, we restricted it for our model, i.e., a system with two layers of players, as follows. Given an unweighted directed graph $G = (V, E)$ containing a vertex set $V$ with $n = |V|$ number of vertices and an edge set $E$ with $M = |E|$ number of edges, the local reaching centrality $c_{R}(i)$ of node $i$ was defined as the ratio of the number of reachable nodes of $i$ through its out-edges, to the total number of nodes that a node could potentially reach (assuming no self-loop). That is,
$$c_{R}(i) = \frac{|S_{i}|}{n-1},$$
where $S_{i} = \{j \in V | 0< d^{out}(i,j) < \infty\}$ is the set of nodes that have nonzero but finite out-distances from node $i$. And the general reaching centrality $GRC(G)$ of  graph $G$ is the normalized sum of all nodes' local reaching centrality by the maximum local reaching centrality:  
\begin{equation}
\label{eq_grc}
GRC(G) = \frac{1}{n-1} \sum_{i \in V} \big[ c^{max}_{R} - c_{R}(i)\big],
\end{equation}
where $c^{max}_{R}$ denotes the largest local reaching centrality in the network. In other words, when $GRC = 1$ and the structure is the most hierarchical, the graph would have only one node--the leader--with nonzero local reaching centrality to all others when others have no out-edge at all. The structure, therefore, represents the (apparent) hierarchy in Mark's model with the highest possible hierarchical score.

Here, to keep generalizing the GRC to two-level groups of nodes, we consider that the players at the top level, however many, have control over others at the bottom and, therefore, each at top has an out-edge toward every other at the bottom. The structures studied in the present research thus become networks, and we define the hierarchicalness $H_{n}(x)$ of this kind of two-level cooperation structures with group size $n$ and $x$ players at the top level using the GRC of network $G$. That is, 
\begin{equation}
\label{eq_hnx}
  H_{n}(x) =
    \begin{cases}
      \frac{(n-x) \cdot \frac{n-x}{n-1}}{n-1} = \big( \frac{n-x}{n-1} \big)^2 & \text{if } 0 < x \leqslant n\\
      0 & \text{if } x = 0.
    \end{cases}       
\end{equation}

\begin{figure}
 \centering
    \includegraphics[width=1\textwidth]{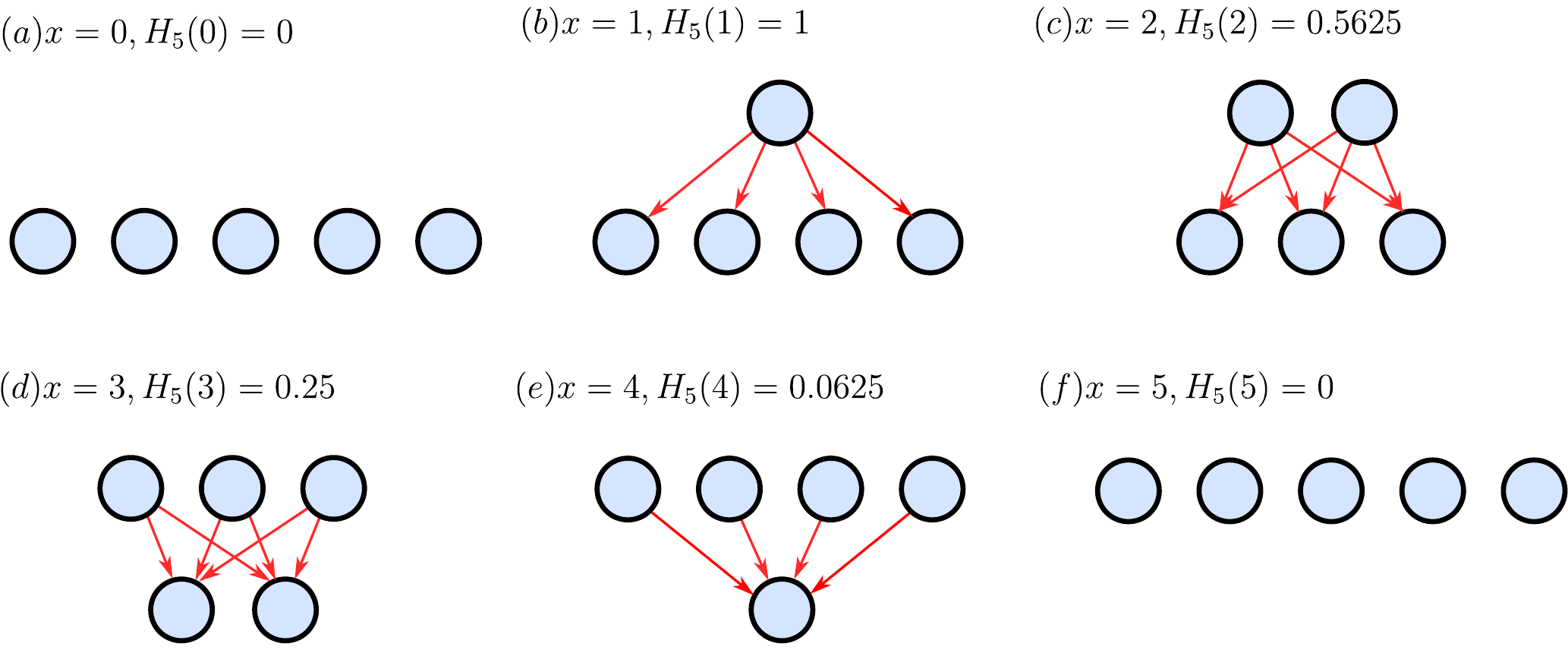}
    \caption{Illustration of $H_{5}(x)$, $x = 0, 1, \cdots 5$.}
    \label{fig_hnx}
\end{figure}

\begin{figure}[!ht]
 \centering
    \includegraphics[width=1\textwidth]{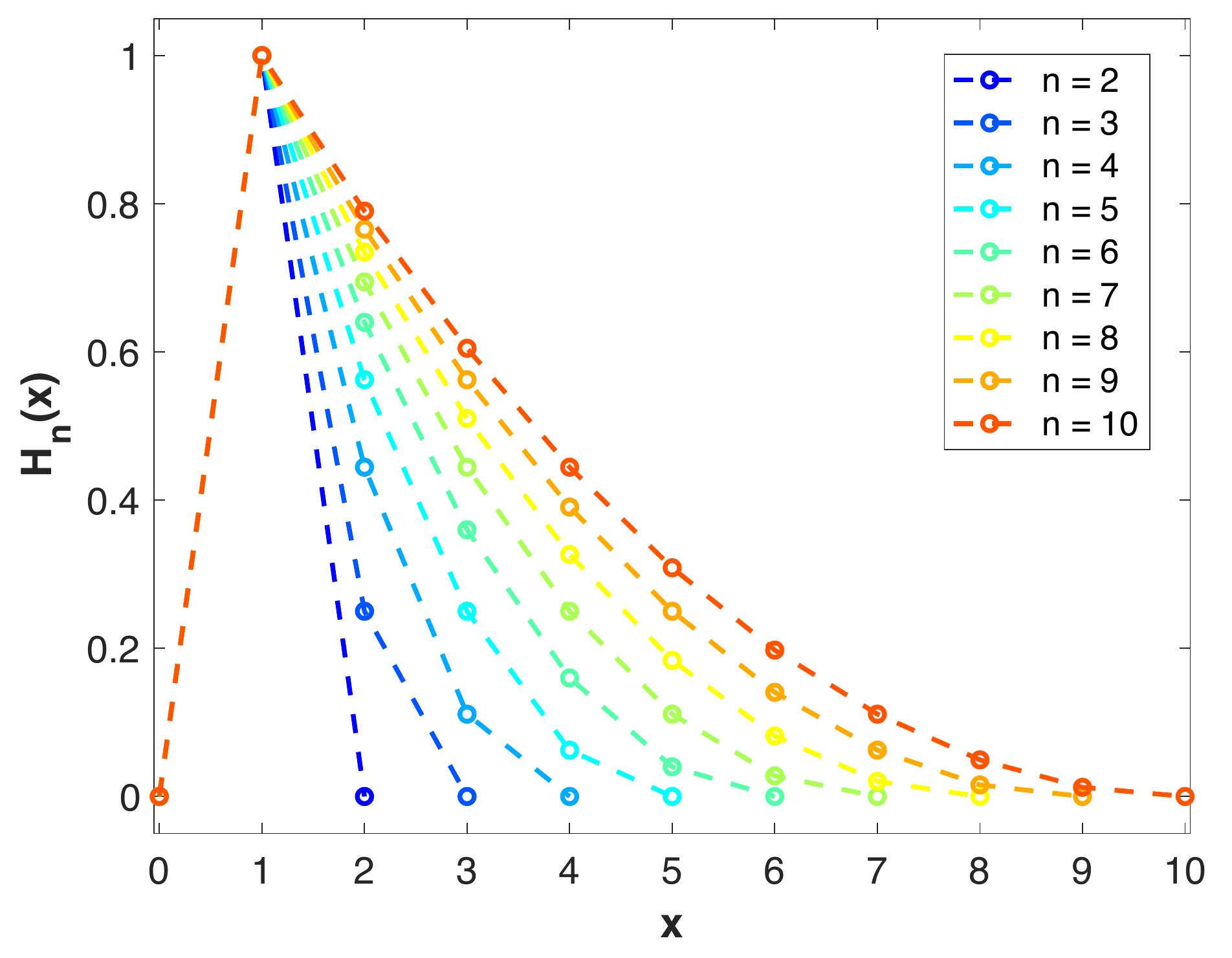}
    \caption{Values of $H_{n}(x)$, for $n = 0, 1, \cdots 10$.}
    \label{fig_hnx_n10}
\end{figure}

Notice here that, when $0 < x \leqslant n$, eq. (\ref{eq_hnx}) applies eq. (\ref{eq_grc}) to a two-level network with $x$ players at the top level. In this case, $c^{max}_{R} = \frac{n-x}{n-1}$ because players at the top show maximum reachability to $n-x$ nodes at the bottom, who do not have this reachability. Moreover, when $x = 0$ and everyone is at the bottom, we set $H_{n}(0) = 0$ to represent the lack of leadership in the group. Readers may consult Figure \ref{fig_hnx} for an illustration of all group structures and their corresponding $H$ values that a group with size $n = 5$  can have. Specifically, in the two extreme cases in the figure where there is no player at the top (subplot (a)) or every node is at the top (subplot (f)), $H = 0$. These two cases demonstrate that, when every node is equal in status, there is zero hierarchy/hierarchicalness in the network. In addition, notice that when $x = 1$, the $H$ value becomes 1, the highest among all scenarios. This thus represents the clear hierarchical structure formed in Mark's EVG. Finally and critically, except for the special case of $x = 0$, for which we stipulate its $H = 0$, $H_{n}(x)$ is a decreasing function of $x$, as required by our conceptual analysis above. Put differently, this decreasing trend suffices that the more leaders at the top level ($x$), the less apparently hierarchical the commanding system ($H$). Readers may also refer to Figure \ref{fig_hnx_n10}, which shows the $H_{n}(x)$ functions with different cases of $n$ and $x$. For these cases, $x$ is defined to be less or equal to the group size $n$, and it can be seen that, again, $H_{n}(x)$ always takes the value between 0 and 1.

Given its useful mathematical properties, $H_{n}(x)$ is used as the proxy to the probability that a status cooperator would contribute to the group. By so doing, as explained above, we make Mark's model a special case of ours, setting the most and the least hierarchical extreme scenarios of our model. Specifically, given a group of size $n$ and $x$ players at the top level, let $H_{n}(x)$ also be the probability of a status cooperator to contribute in any group structure emerging from the prepay communication. As such, if there is only one player signaling their interest in higher status and, thus, rise to the top, all status cooperators will contribute, that is, with a probability $H_{n}(1) = 1$, as in Mark's model. On the other hand, if there is no player at the top level, all status cooperators will not contribute, that is, with a probability $H_{n}(0) = 0$, again as in Mark's model. Together, these two special cases then cover the cases Mark discusses. Beyond these, the group structure in our model can also have any number of leaders, so we are able to delineate the cooperation process wherein there is any arbitrary number of players in the leader role.

\clearpage

\clearpage

\subsection{Random mixing}

As mentioned, we follow the overall settings of Mark's original game and only modify the process of preplay communication and the corresponding contribution of status cooperators ($SC$ in Mark's paper, now we use $C$). The determination of groups and the population dynamics remain the same along with the evolution. Therefore, our changes to Mark's model should have merely affected its expected payoffs for the status cooperator $W(C)$ and the defector $W(D)$ within one iteration. Everything in-between iterations holds. As such, below, we focus on $W(C)$ and $W(D)$ first, and then analyze the corresponding changes in predicted cooperation as a result of changed expected payoffs.

To illustrate the changes in $W(C)$ and $W(D)$, here we start from a small group of size $n$, assuming an unbiased random mixing among cooperators and defectors. When $n = 2$, the expressions of the expected payoffs for the status cooperators and defectors are shown in formula (\ref{wc_ours_n2}) and (\ref{wd_ours_n2}), respectively. With the probability for our protagonist, the status cooperator, to meet another cooperator being $f_{c}$ and to meet a defector being $1- f_{c}$, in equation (\ref{wc_ours_n2}), the first and the second line are the expected payoff for a status cooperator meeting another cooperator and a defector, respectively. Particularly, there are three terms in the first line of equation (\ref{wc_ours_n2}); they in turn represent the situations where the focal cooperator meets another cooperator while there are 2, 0, or 1 player at the top level. As a result, when all players (i.e., 2) or no player is leaders, no status cooperator contributes, and each of them takes $c$ back. By contrast, when there is 1 and only leader in the group--the special case of $H_{n}(1)$--every status cooperator contributes equally and receives $b$ equally. Moreover, the two terms of the second line, in turn, represent the case wherein the status cooperator meets a defector, does not go to the top level themself, and hence not contribute, and the case wherein the cooperator goes to the top and contributes. Notice that in the latter scenario, the contribution $b$ will be shared by the two players in the group since the defector does not contribute.
\begin{equation}
\begin{aligned}
\label{wc_ours_n2}
W(C) &= f_{c} \cdot \bigg\{ {2 \choose 2} \bigg( \frac{1}{2}\bigg)^{2} \bigg( \frac{2-1}{2}\bigg)^{0} \cdot c + {2 \choose 0} \bigg( \frac{1}{2}\bigg)^{0} \bigg( \frac{2-1}{2}\bigg)^{2} \cdot c + {2 \choose 1} \bigg( \frac{1}{2}\bigg)^{1} \bigg( \frac{2-1}{2}\bigg)^{1} \cdot b \bigg \} \\
& + (1-f_{c}) \cdot \bigg\{{1 \choose 0} \bigg( \frac{1}{2}\bigg)^{0} \bigg( \frac{2-1}{2}\bigg)^{1} \cdot c + {1 \choose 1} \bigg( \frac{1}{2}\bigg)^{1} \bigg( \frac{2-1}{2}\bigg)^{0} \cdot \frac{b}{2} \bigg \}
\end{aligned}
\end{equation}

Turning to the expected payoff for the defector, here we express it in equation (\ref{wd_ours_n2}) in that, whether the defector meets a status cooperator or another defector, he always keeps his cost $c$--the first term in the equation. Therefore we only need to focus on whether they can get additional shares of $b$. In the current case of $n = 2$, the defector would only obtain the benefit $b$ when meeting a cooperator who goes to the top level (therefore contributes). This is the second term in the line.

\begin{equation}
\begin{aligned}
\label{wd_ours_n2}
W(D) = c + f_{c}  \cdot \bigg\{ {1 \choose 1} \bigg( \frac{1}{2}\bigg)^{1} \bigg( \frac{2-1}{2}\bigg)^{0} \cdot \frac{b}{2} \bigg\} 
\end{aligned}
\end{equation}

Now, generalizing the equations, the hierarchicalness function $H_{n}(x)$ joins in to form the payoffs for a larger group size $n$. Notice that $x = 1$ is always a special case. Here, the status hierarchy is apparent and all cooperators contribute. The other special case is that all status cooperators are at the same level, either at the top or the bottom. In this case, no cooperator contributes. Accordingly, below are the the expected payoffs for the cooperator and the defector when $n = 3$, in equation (\ref{wc_ours_n3}) and (\ref{wd_ours_n3}), respectively.

\begin{equation}
\begin{aligned}
\label{wc_ours_n3}
W(C) & = (1-f_{c})^{2} \cdot \bigg( \frac{2}{3} \bigg)\cdot c\\
& + {2 \choose 1}  f_{c} (1-f_{c}) \cdot \bigg \{ \bigg( \frac{2}{3} \bigg)^2 + \bigg( \frac{1}{3} \bigg)^2 (1-H_{3}(2))\bigg\} \cdot c\\
& + (f_{c})^2 \cdot \bigg \{ \bigg( \frac{2}{3} \bigg)^3 + {3 \choose 2} \bigg( \frac{1}{3} \bigg)^2 \bigg( \frac{2}{3} \bigg)^2 (1-H_{3}(2)) + \bigg( \frac{1}{3} \bigg)^3 \bigg\} \cdot c\\
& + (1-f_{c})^{2} \cdot \bigg( \frac{1}{3} \bigg) \bigg( \frac{1}{3} \bigg) \cdot b \\
& + {2 \choose 1}  f_{c} (1-f_{c}) \cdot \bigg \{ {2 \choose 1} \bigg( \frac{1}{3} \bigg) \bigg( \frac{2}{3} \bigg) \bigg( \frac{2}{3} \bigg) \\
& + {2 \choose 2} \bigg( \frac{1}{3} \bigg)^2 \cdot \bigg[ 
   {2 \choose 1}(H_{3}(2))(1-H_{3}(2))\bigg( \frac{1}{3} \bigg)
+ {2 \choose 2}(H_{3}(2))^{2}\bigg( \frac{2}{3} \bigg)
\bigg] \bigg\} \cdot b \\
& + (f_{c})^2 \cdot \bigg\{ {3 \choose 1} \bigg( \frac{1}{3} \bigg) \bigg( \frac{2}{3} \bigg)^2 + 
{3 \choose 2} \bigg( \frac{1}{3} \bigg)^2 \bigg( \frac{2}{3} \bigg)^1\\
& \cdot \bigg[ 
{3 \choose 1} (H_{3}(2))(1-H_{3}(2))^2 \bigg( \frac{1}{3} \bigg) + 
{3 \choose 2} (H_{3}(2))^2(1-H_{3}(2))^1 \bigg( \frac{2}{3} \bigg) +
(H_{3}(2))^3 \bigg] \bigg\} \cdot b
\end{aligned}
\end{equation}

\begin{equation}
\begin{aligned}
\label{wd_ours_n3}
W(D) &= c + 2f_{c}(1-f_{c}) \cdot \bigg\{ \bigg( \frac{1}{3}\bigg) (H_{3}(1))  \bigg( \frac{1}{3}\bigg) \bigg\} \cdot b \\
& + f^{2}_{c} \bigg\{ {2 \choose 1} \bigg( \frac{1}{3}\bigg) \bigg( \frac{2}{3}\bigg) \bigg( \frac{2}{3} \bigg) + \bigg( \frac{1}{3}\bigg)^{2} 
\bigg[ {2 \choose 1} (H_{3}(2)) (1-H_{3}(2)) \bigg( \frac{1}{3}\bigg)+ (H_{3}(2))^{2} \bigg( \frac{2}{3}\bigg) \bigg] \bigg\} \cdot b
\end{aligned}
\end{equation}

The first three lines in equation (\ref{wc_ours_n3}) represent the situation in which a status cooperator doesn't contribute and take back its cost $c$, lines 4 to 8 represent the situation in which the cooperator can earn the benefit $b$. In each scenario, we need to consider the focal status cooperator meets two defectors (with probability $(1-f_{c})^2$), one cooperator and one defector (with probability ${2 \choose 1}f_{c}(1-f_{c})$), and two cooperators (with probability $f^2_{c}$).  Here, we focus on explaining the last two lines, where the status cooperator meets two cooperators as the rest is obtained following the same calculation. The status cooperator can always take back $b$ if there is one cooperator at the top level. Assuming there are two cooperators at the top level (i.e., $x = 2$), though a cooperator to start, this player acts as a defector in the end, given the probability $1- H_{3}(2)$ ``not" to follow the lead. This probability is always 0 in Mark's model because of $1- H_{3}(1) = 0$, but does not have to be so in ours. Here, one needs to consider two different scenarios: If the focal cooperation does not contribute, they can keep $c$ and obtain extra portions of $b$ depending on the number of players who contribute. Consequently, the focal player can take $\frac{k}{3}$ shares of $b$ if there are $k$ cooperators who end up contributing.

On the other hand, equation (\ref{wd_ours_n3}) is the expected payoff for the defector when $n = 3$. Note that, again, a defector can always take back $c$. They then obtains additional fractions of $b$ if they meets one status cooperator--with probability $2f_{c}(1-f_{c})$ (the second term in the first line) and two cooperators--with probability $f^{2}_{c}$ (the last line). Specifically, the first term inside the part following $f^{2}_{c}$ represents that exactly one status cooperator becomes the leader, so both status cooperators contribute. The second term in line 2 represents the scenario where both status cooperators become leaders, and therefore, each of them has a probability $H_{3}(2)$ to contribute. In this case, the first term of in the square bracket expresses that one cooperator contributes and the other does not, and the second term shows both of them do. As a result, the focal defector obtains $\frac{b}{3}$ and $\frac{2b}{3}$ of benefits, respectively. The last line then follows the same rationale as in the prior analysis.

Finally, the expected payoffs for a status cooperator and a defector in a general case of arbitrary $n$ are identified in equations (\ref{wc_ours}) and (\ref{wd_ours}), respectively. To construct the equations, we aggregated the payoff that a focal player obtains based on three quantities: (i) the number of status cooperators in the group, (ii) the number of status cooperators who become leaders, and (iii) the number of status cooperators who end up contributing to the group. Together, the expected payoff of the player can be expressed with a triple summation of the three quantities, denoted with dummy variables $i$, $j$, and $k$, corresponding to the quantities of (i), (ii), and (iii). Here, it might be worth noting that (ii) depends on (i) and (iii) depends on them both. Consequently, the first two lines of equation (\ref{wc_ours}) describe the scenarios in which the focal status cooperator does not contribute and takes back $c$, whereas the last two lines describe the scenarios in which they take portions of $b$ depending on quantities (i), (ii), and (iii). On the other hand, equation (\ref{wd_ours}) indicates that the defector always keeps $c$ with extra shares of $b$ similarly depending on (i), (ii), and (iii).

\begin{align}
   \begin{split}
\label{wc_ours}
W(C) &= \sum^{n-1}_{i=0} {n-1 \choose i} (f_{c})^{i} (1-f_{c})^{n-1-i} \cdot
 \bigg[ \sum^{i+1}_{j = 0} {i+1 \choose j} \bigg(\frac{1}{n} \bigg)^{j} \bigg( \frac{n-1}{n}\bigg)^{i+1-j} \big(1- H_{n}(j)\big) \bigg]  \cdot c\\
&+  \sum^{n-1}_{i =0} {n-1 \choose i} (f_{c})^{i} (1-f_{c})^{n-1-i} \cdot \bigg \{ 
 \sum^{i+1}_{j = 0} {i+1 \choose j} \bigg(\frac{1}{n} \bigg)^{j} \bigg( \frac{n-1}{n}\bigg)^{i+1-j} \\
& \cdot \bigg[ \sum^{i+1}_{k = 0} {i+1 \choose k}\big( H_{n}(j)\big)^{k} \big(1- H_{n}(j)\big)^{i+1-k}  \bigg( \frac{k}{n}\bigg)   \bigg] \bigg\} \cdot b
   \end{split}
\end{align}

\begin{equation}
\begin{aligned}
\label{wd_ours}
W(D) &=   c +\sum^{n-1}_{i =0} {n-1 \choose i} (f_{c})^{i} (1-f_{c})^{n-1-i} \\
& \cdot \bigg\{ \sum^{i}_{j = 0} {i \choose j} \bigg(\frac{1}{n} \bigg)^{j} \bigg( \frac{n-1}{n}\bigg)^{i-j}
\cdot \bigg[ \sum^{i}_{k = 0} {i \choose k}\big( H_{n}(j)\big)^{k} \big(1- H_{n}(j)\big)^{i-k}  \bigg( \frac{k}{n}\bigg)   \bigg] \bigg\} \cdot b
\end{aligned}
\end{equation}

Validating the general equations (\ref{wc_ours}) and (\ref{wd_ours}), we further built an agent-based simulation (see Appendix for the code) of our EVG. As shown in Figure \ref{comparison_random}, we compared the analytical and the simulated predictions of the internal equilibrium--i.e., the cost-to-benefit ratio $c/b$ that makes $W(C) = W(D)$--of groups with different sizes $n$ and fractions of status cooperators $f_{c}$. A dot in the figure is an average of 100,000 simulations of a given pair of $n$ and $f_{c}$, and the lines are the analytical predictions of their same-color dots. As can be seen, the equations that based the predicted lines--equations (\ref{wc_ours}) and (\ref{wd_ours})--yielded nearly perfect matches with the simulations, bolstering the validity of our formal model. In addition, note that above a line, the cost becomes higher and $W(D) > W(C)$. Therefore status cooperators will decrease in number in the next iteration of the population dynamics. By contrast, below a line, the cost is lower and $W(D) < W(C)$. Therefore cooperators will increase in the next iteration.

\begin{figure}[!ht]
 \centering
    \includegraphics[width=1\textwidth]{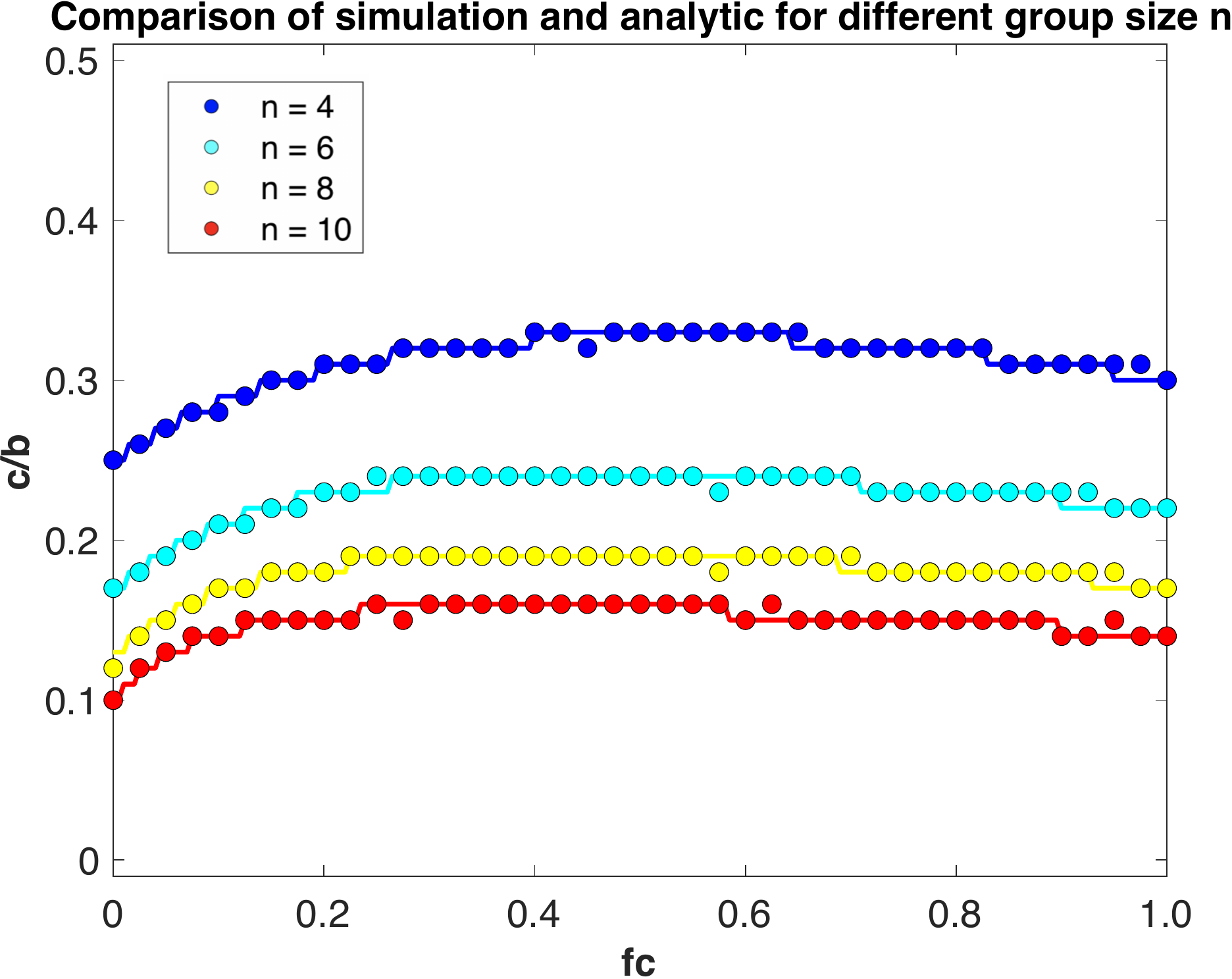}
    \caption{Comparison of simulations and analytics for different group size $n$. Dots and lines represent simulation and analytical results respectively.}
    \label{comparison_random}
\end{figure}

Having established the model, subsequently, this section examines our main research question about the evolutionary emergence of cooperation against defection in hierarchical groups that are allowed to have multiple leaders. As does Mark, we first present results of the analysis of the cost-to-benefit range wherein status cooperation is a stable evolutionary strategy in social dilemmas. To determine the range, we searched for the upper and the lower bound for cooperation to remain stable, following the reasoning in Mark's paper. Specifically, our multi-leader model would not change the lower bound that Mark obtained, because only one status cooperator would be placed in a group without any other cooperator and multi-leader structure is simply impossible.  This made the lower bound of the cost-to-benefit ratio $\frac{1}{n}$ as in Mark's work. Specifically, if the ratio is smaller than $\frac{1}{n}$, $b$ becomes too large relative to $c$ to keep the game a dilemma: Cooperation will always outperform defection regardless of whom cooperators interact with and how the two interact. 

Moreover, to investigate the upper bound of the cost-to-benefit range that makes cooperation a stable strategy in a social dilemma--that is, the conditions under which that $f_{c} = 1$ is stable--we again followed the approach taken by Mark. That is, we considered a population of status cooperators and then replace one of them with a defector. If these cooperators' expected payoff in the original homogeneous group is larger than that for the invading defector, it is said that cooperation is an evolutionarily stronger and hence stable strategy against defection. Conversely, if it is expected that payoffs accrue faster for the invading defector than for the group of cooperators, the defector will eventually take over the group. Mathematically, this conceptual analysis means to first compute a status cooperator's payoff under $f_{c} = 1$. If it is larger than a defector's payoff under $f_{c} = \frac{n-1}{n}$, then cooperation is stable. Accordingly, we plugged these condition in equations (\ref{wc_ours}) and (\ref{wd_ours}) to find the boundary of $c/b$. 

\begin{figure}[!ht]
 \centering
    \includegraphics[width=1\textwidth]{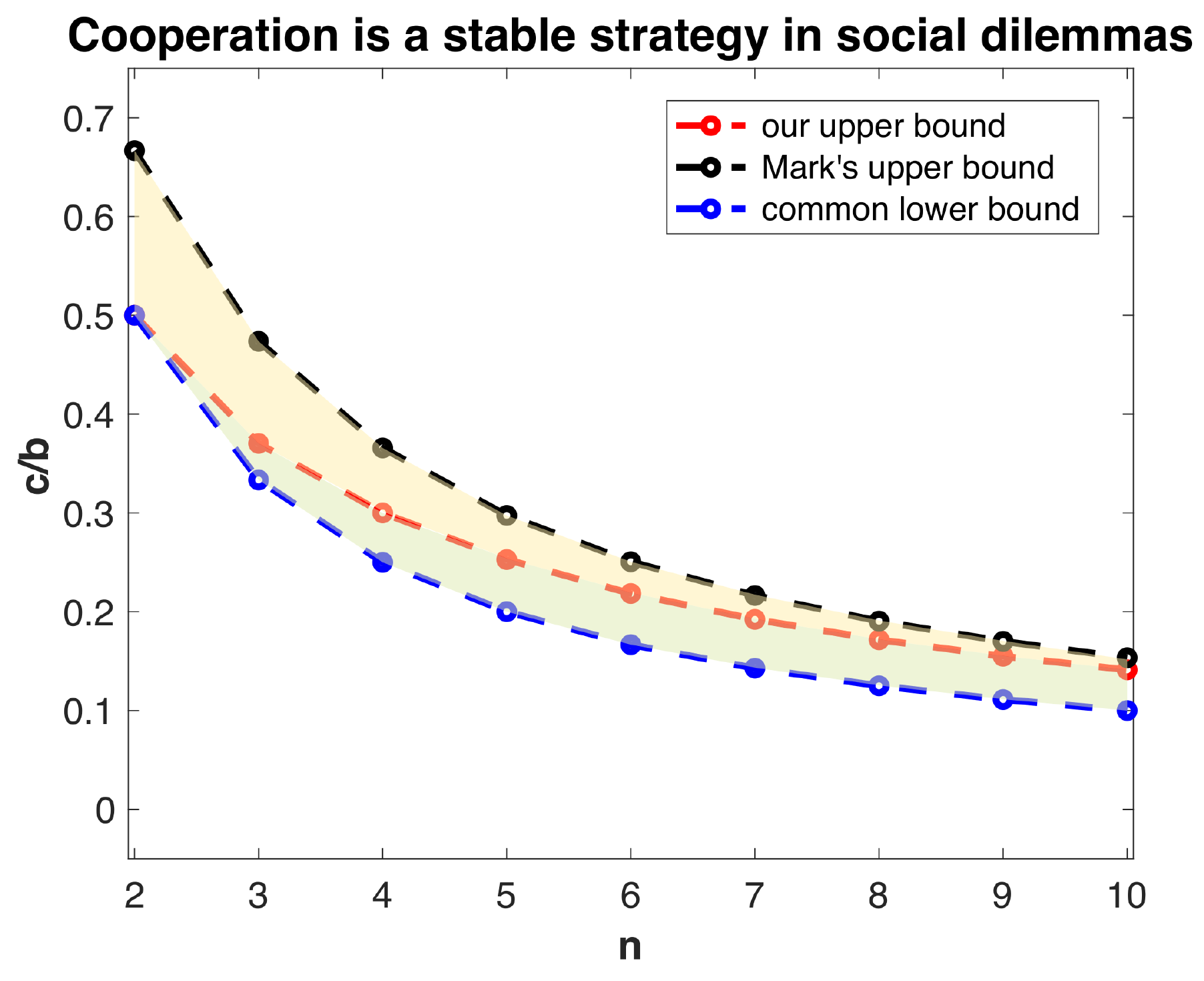}
    \caption{Cooperation is a stable strategy in social dilemmas. Our model and Mark's model have the same lower bound. The upper bound in our model (shown in red dots and lines) is larger than in Mark's (shown in black dots and lines).}
    \label{social_dilemma}
\end{figure}

\begin{figure}[!ht]
 \centering
    \includegraphics[width=1\textwidth]{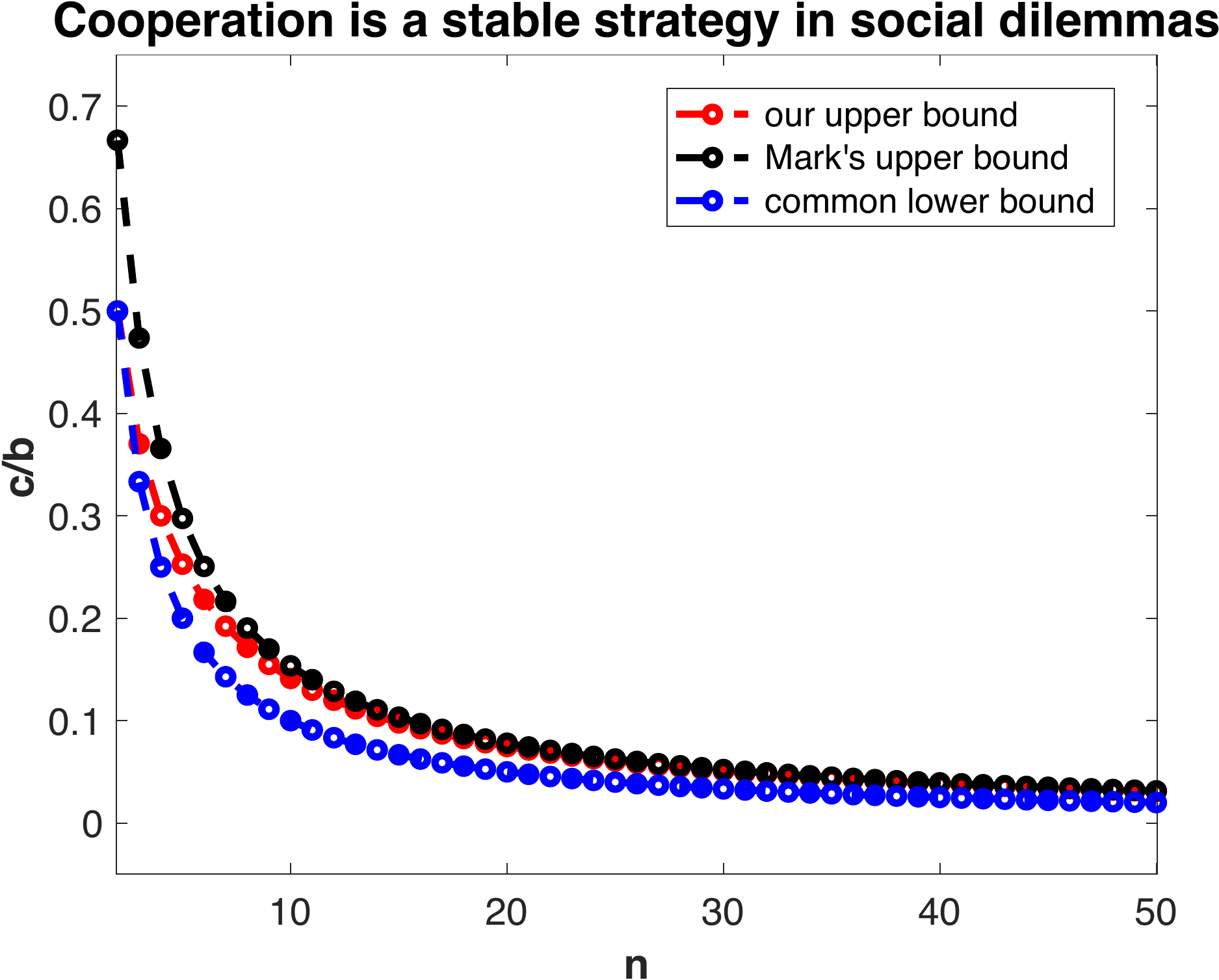}
    \caption{Cooperation is a stable strategy in social dilemmas for large $n$. Our model and Mark's model have the same lower bound. The upper bound in our model (shown in red dots and lines) is larger than in Mark's (shown in black dots and lines).}
    \label{social_dilemma_n50}
\end{figure}

As shown in Figure \ref{social_dilemma}, although sharing the same lower bound, Mark's and our model do not have the same upper bound: Our upper bound of $c/b$--derived numerically using equations (\ref{wc_ours}) and (\ref{wd_ours})--is lower and, the difference between our upper bound and Mark's gets smaller as the group size $n$ increases. Since the existence of multi-leadership in our model reduces the probability to cooperate, the region of the cost-to-benefit ratio in which status cooperation is a stable strategy in social dilemmas is narrower when multi-leader hierarchy is permitted than when it is not, and this is specifically caused by reduced difficulty for defectors to invade cooperators, as opposed to the heightened difficulty for cooperators to invade defectors. We also examine the region of $c/b$ in larger groups, because it can be derived analytically that the upper bound of Mark's model will approach 0 and, thus, eliminate the region all together as $n$ goes to infinity. As shown in Figure \ref{social_dilemma_n50}, even though we also found that the upper bound of our model seemingly approaches 0 as $n$ goes to infinity as does Mark's, the stability region of cooperation under our multi-leader EVG still seems wider than that under Mark's single-leader EVG. 

\begin{figure}[!ht]
 \centering
    \includegraphics[width=1\textwidth]{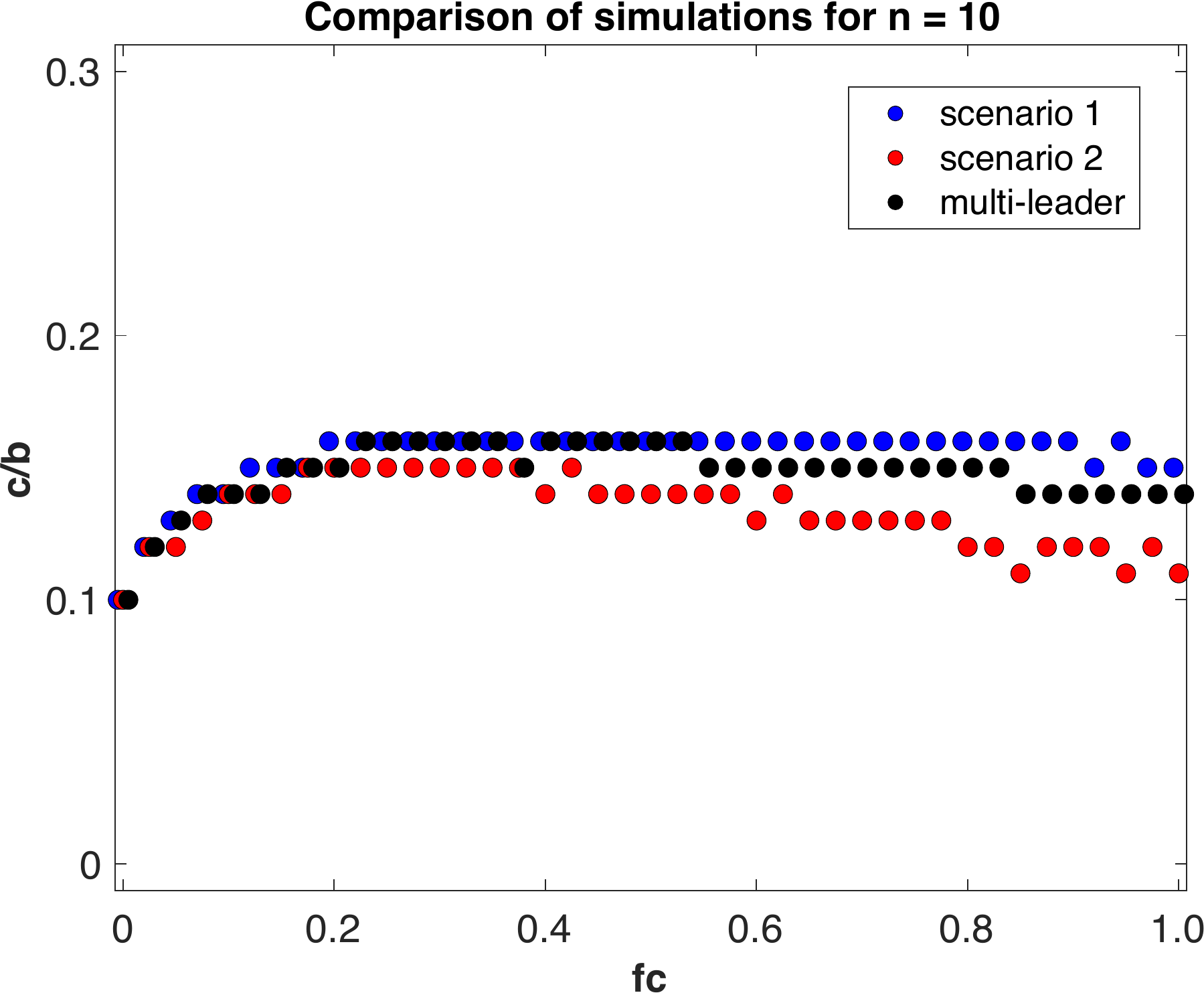}
    \caption{Comparison of simulations of the boundary $W(C) = W(D)$ as $f_{c}$ increases when $n = 10$.}
    \label{figX}
\end{figure}

Having derived that cooperation can be stable and, thus, may emerge in evolution, we now turn to a critical factor that Mark reports affects the emergence of cooperation: the cooperator-to-defector ratio in groups $f_{c}$. In Figure \ref{figX}, using simulated results, we plot the cost-to-benefit ratio $c/b$ that makes $W(C) = W(D)$, against $f_{c}$ when an arbitrary $n = 10$, to illustrate how $f_{c}$ changes the dynamics and relative strength of status cooperation and defection. Each dot is an average of 100,000 simulations. Again, as in Figure \ref{comparison_random}, the region above dots of the same color represents the space where defectors are expected to outperform cooperators in a round of the game, whereas the region below represents the space where the situation reverses. Further, to compare these dynamics in single-leader groups to that in multi-leader ones, we plot the dots of our EVG as well as Mark's. Here, it might be worth noting that there are two but not one line of dots for Mark's model. It is because we see two slightly different scenarios from Mark's reasoning; both are interesting and useful in gauging the effects of multi-leader structures. In particular, the game could also be set up so that, whenever there are multiple cooperators signaling willingness to take the leader role in the initial round of leader selection, showing the group's preference for a status difference, the selection repeats until the wish is realized and a leader is agreed on (scenario 1). By contrast, a single-leader game could be set up in the way that whenever there is more than one cooperator nominated to be the leader in a round of nomination, the nomination starts over entirely and the process may end up producing no leader at all (scenario 2). That is, once begun, the process does not fall back to the no-leader case, and this is what Mark's model equation implies.

Combing our model with the two potential interpretations of Mark's work, Figure \ref{figX} then reveals some novel effects of multi-leader hierarchy. In the figure, it can be seen that the equilibrium state $W(C) = W(D)$ in which multi-leader structures are permitted sits in between the situations where they are not permitted and simply forgot and where they are not directly used but still help guarantee a later single-leader structure. Here, for the comparison between our model and the model  ``with memory," it is anticipated that our multi-leader game relatively suppresses cooperation to defection--its line of dots is lower--because here the single-leader game can be effectively thought of as a multi-leader game with no discount on the probability to cooperate. In other words, whenever there are cooperators nominated in the initial round of selection, all cooperators contribute with no hesitation later. Accordingly, the more cooperators in a group, the more likely one of them will show the interest to lead and then build a fully collaborative team of cooperators. On the other hand, not only may the likelihood to contribute be discounted in our EVG; the more cooperators are there, the more likely a multi-leader group will surface and reduce group contribution. 

What is intriguing here, consequently, is the finding that cooperation in our EVG outperforms that in a game in which multi-leader hierarchy is simply ignored--the ``no-memory" model. Because of insisting on reaching a clear consensus about leadership, this kind of single-leader model would require all multi-leader structures to rearrange, in the fact that the selection of leaders may well generate no status hierarchy in the end and make everyone a defector. From this perspective, allowing attenuated parallel leadership is then a worthy compromise. Although the full contribution is not extracted there, at least some of it is, and this is better than nothing. For theory, the present paper therefore elaborates on Mark's investigation, and documents not only why hierarchy may exist, but also why its leadership is sometimes distributed to multiple: It may be worth it.

Finally, focusing on our multi-leader EVG, Figure \ref{figX} suggests that there may be an optimal level of $f_{c}$; it is not the higher the better for the ration in terms of bolstering cooperation, not as Mark's equation (i.e., the model with memory) predicts. In Figures X as well as Figure \ref{comparison_random}, we report that even if the $W(C) = W(D)$ lines of dots are in general in an ascending trend as $f_{c}$ increases, all lines of dots slightly yet consistently drop lower when $f_{c}$ approaches 1. The phenomenon echos with our above reasoning that, when there are too many cooperators competing for leadership in a group, forcing the group to choose more leaders and compromise (collaborative) members' willingness to contribute, cooperators in effect become partial defectors, and that creates more harm than good. To put it differently, even if there are few defectors in a group, cooperators may still break the team among themselves, as can be seen in everyday experiences \citep{chrobot2009challenge}. This result of the current study consequently extends Mark's work and helps shed light one why not everyone in this world today behaves collectively and collaboratively and why that may in fact make the society function better.

\clearpage

\subsection{Assortative mixing}

Following closely with Mark, in this section, we keep generalizing our model to explore how it behaves under non-random, assortative mixing. In Mark's original paper, he reports that a status cooperator may be able to reversely invade a population of defectors with the help of enough assortativity, because assortative mixing unequally protects and, therefore, favors cooperative strategies. Following this analysis, we adopted the assortative mixing rules developed in Mark's paper, that is, the selection of the first individual of a group is unbiased and random yet the successive members of the group are biasedly selected depending on the assortativity parameter $\tau$. To construct this analysis in our model, the only change to make lay in the rate of meeting a cooperator $f_{c}$. We thus derived the expected payoffs for a status cooperator and a defector in a multi-leader model with assortative mixing in equations (\ref{wc_tau}) and (\ref{wd_tau}) as follows.

\begin{equation}
\begin{aligned}
\label{wc_tau}
W(C) &=  \sum^{n-1}_{i=0} \Bigg\{ \bigg(\frac{i+1}{n} \bigg) \bigg\{ \bigg(\frac{1}{i+1} \bigg) {n-1 \choose i} \big[\tau + (1-\tau)f_{c}\big]^{i} \big[(1-\tau)(1-f_{c})\big]^{n-i-1}\\
&+\bigg(\frac{i}{i+1}\bigg) f_{c} {n-1 \choose i} \big[\tau+(1-\tau)f_{c}\big]^{i-1} \big[(1-\tau)(1-f_{c})\big]^{n-i-1}\bigg\}  \\
&+ \bigg(\frac{n-i-1}{n}\bigg) \bigg\{ (1-f_{c}) {n-1 \choose i} \big[(1-\tau)f_{c}\big]^{i} \big[\tau + (1-\tau)(1-f_{c})\big]^{n-i-2} \bigg\} \Bigg\}\\
& \cdot
 \bigg[ \sum^{i+1}_{j = 0} {i+1 \choose j} \bigg(\frac{1}{n} \bigg)^{j} \bigg( \frac{n-1}{n}\bigg)^{i+1-j} \big(1- H_{n}(j)\big) \bigg]  \cdot c\\
&+ \sum^{n-1}_{i=0} \Bigg\{ \bigg(\frac{i+1}{n} \bigg) \bigg\{ \bigg(\frac{1}{i+1} \bigg) {n-1 \choose i} \big[\tau + (1-\tau)f_{c}\big]^{i} \big[(1-\tau)(1-f_{c})\big]^{n-i-1}\\
&+\bigg(\frac{i}{i+1}\bigg) f_{c} {n-1 \choose i} \big[\tau+(1-\tau)f_{c}\big]^{i-1} \big[(1-\tau)(1-f_{c})\big]^{n-i-1}\bigg\}  \\
&+ \bigg(\frac{n-i-1}{n}\bigg) \bigg\{ (1-f_{c}) {n-1 \choose i} \big[(1-\tau)f_{c}\big]^{i} \big[\tau + (1-\tau)(1-f_{c})\big]^{n-i-2} \bigg\} \Bigg\}\\
& \cdot \bigg \{ 
 \sum^{i+1}_{j = 0} {i+1 \choose j} \bigg(\frac{1}{n} \bigg)^{j} \bigg( \frac{n-1}{n}\bigg)^{i+1-j} \\
& \cdot \bigg[ \sum^{i+1}_{k = 0} {i+1 \choose k}\big( H_{n}(j)\big)^{k} \big(1- H_{n}(j)\big)^{i+1-k}  \bigg( \frac{k}{n}\bigg)   \bigg] \bigg\} \cdot b
\end{aligned}
\end{equation}

\begin{equation}
\begin{aligned}
\label{wd_tau}
W(D) &=   c +  \sum^{n-1}_{i=0} \Bigg\{ \bigg(\frac{i}{n} \bigg) \bigg\{ f_{c} {n-1 \choose i} \big[\tau + (1-\tau)f_{c}\big]^{i-1} \big[(1-\tau)(1-f_{c})\big]^{n-i-1}\bigg\} \\
&+\bigg(\frac{n-i}{n}\bigg) \bigg\{ \bigg(\frac{1}{n-i} \bigg) {n-1 \choose i} \big[(1-\tau)f_{c}\big]^{i} \big[\tau + (1-\tau)(1-f_{c})\big]^{n-i-1} \\
&+ \bigg(\frac{n-i-1}{n-i}\bigg) (1-f_{c}) {n-1 \choose i} \big[(1-\tau)f_{c}\big]^{i} \big[\tau + (1-\tau)(1-f_{c})\big]^{n-i-2}      \bigg\}\Bigg\} \\
&\cdot \bigg\{  \sum^{i}_{j = 0} {i \choose j} \bigg(\frac{1}{n} \bigg)^{j} \bigg( \frac{n-1}{n}\bigg)^{i-j} \cdot \bigg[ \sum^{i}_{k = 0} {i \choose k}\big( H_{n}(j)\big)^{k} \big(1- H_{n}(j)\big)^{i-k}  \bigg( \frac{k}{n}\bigg)   \bigg] \bigg\} \cdot b
\end{aligned}
\end{equation}

Equations (\ref{wc_tau}) and (\ref{wd_tau}) may look complex, yet they are merely incremental from the general random-mixing equations (\ref{wc_ours}) and (\ref{wd_ours}). As mentioned, only terms related to the fraction of status cooperators $f_{c}$ are modified, and there are only two such terms. 

In the first and the second line of equation (\ref{wc_ours}), the term
$$\sum^{n-1}_{i =0} {n-1 \choose i} (f_{c})^{i} (1-f_{c})^{n-1-i}$$
is substituted with 
\begin{equation*}
\begin{aligned}
&\sum^{n-1}_{i=0} \bigg(\frac{i+1}{n} \bigg) \bigg\{ \bigg(\frac{1}{i+1} \bigg) {n-1 \choose i} \big[\tau + (1-\tau)f_{c}\big]^{i} \big[(1-\tau)(1-f_{c})\big]^{n-i-1}\\
&+\bigg(\frac{i}{i+1}\bigg) f_{c} {n-1 \choose i} \big[\tau+(1-\tau)f_{c}\big]^{i-1} \big[(1-\tau)(1-f_{c})\big]^{n-i-1}\bigg\}\\
&+ \bigg(\frac{n-i-1}{n}\bigg) (1-f_{c}) {n-1 \choose i} \big[(1-\tau)f_{c}\big]^{i} \big[\tau + (1-\tau)(1-f_{c})\big]^{n-i-2}.
\end{aligned}
\end{equation*}

Here, the first two lines of the expression represent the scenario wherein the first member of a group is a status cooperator. The term with $\frac{1}{i+1}$ is the probability for the focal cooperator to be the first member and $i$ others to be chosen as status cooperators; the term with $\frac{i}{i+1}$ is the probability for another status cooperator to be the first member and $n-1$ others including the focal player to be cooperators. Lastly, the third line of the expression describes the case in which the first member of a group is a defector and they meet $i+1$ status cooperators including the focal player. 

With aforementioned modifications, equation (\ref{wc_tau}) was then obtained. Importantly, when $\tau = 0$, i.e., random mixing, this equation (\ref{wc_tau}) reduces to equation (\ref{wc_ours}) as expected. Similarly, equation (\ref{wd_tau}) was obtained by changing the terms related to $f_{c}$ in equation (\ref{wd_ours}). The term in first line of (\ref{wd_ours})
$$\sum^{n-1}_{i =0} {n-1 \choose i} (f_{c})^{i} (1-f_{c})^{n-1-i}$$
is substituted with 
\begin{equation*}
\begin{aligned}
&\sum^{n-1}_{i=0} \bigg(\frac{i}{n} \bigg) \bigg\{ f_{c} {n-1 \choose i} \big[\tau + (1-\tau)f_{c}\big]^{i-1} \big[(1-\tau)(1-f_{c})\big]^{n-i-1}\bigg\} \\
&+\bigg(\frac{n-i}{n}\bigg) \bigg\{ \bigg(\frac{1}{n-i} \bigg) {n-1 \choose i} \big[(1-\tau)f_{c}\big]^{i} \big[\tau + (1-\tau)(1-f_{c})\big]^{n-i-1} \\
&+ \bigg(\frac{n-i-1}{n-i}\bigg) (1-f_{c}) {n-1 \choose i} \big[(1-\tau)f_{c}\big]^{i} \big[\tau + (1-\tau)(1-f_{c})\big]^{n-i-2}      \bigg\}.
\end{aligned}
\end{equation*}

The first line of the expression shows the scenario wherein the first member of a group is a cooperator and they meet $i-1$ other status cooperators, and the second and third lines express the scenario in which the first member of a group is a defector. The term with $\frac{1}{n-i}$ is the probability for the defector to be the first member with $i$ others being status cooperators, and the term with $\frac{n-i-1}{n-i}$ is the probability for another defector to be the first member with $i$ other in the group being cooperators. Following these changes, equation (\ref{wd_tau}) was subsequently formed. Again, when $\tau = 0$, this equation (\ref{wd_tau}) reduces to equation (\ref{wd_ours}).

\begin{figure}[!ht]
 \centering
    \includegraphics[width=1\textwidth]{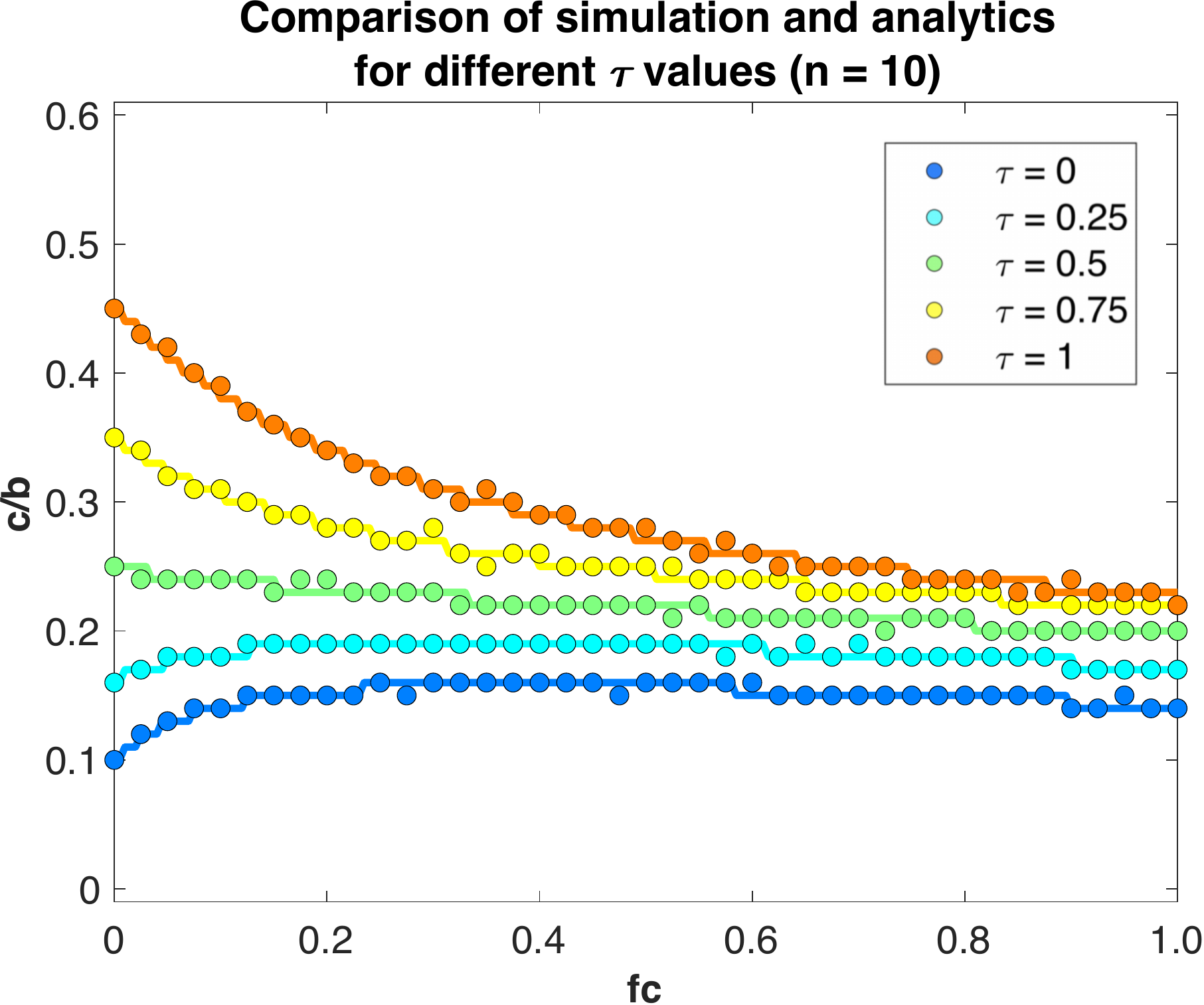}
    \caption{Comparison of simulations and analytics of the lines $W(C) = W(D)$ with different mixing ratio $\tau$. Dots and lines represent simulation and analytical results respectively.}
    \label{tau_comparison}
\end{figure}

As before, to validate equations (\ref{wc_tau}) and (\ref{wd_tau}), we built an agent-based simulation of our multi-leader model, now with the possibility of assortative mixing (see Appendix for the code). Following the same simulation procedure, Figure \ref{tau_comparison} shows the analytics and the simulations of assortatively mixed groups, in their cost-to-benefit ratio $c/b$ against the fraction of status cooperators $f_{c}$ given internal equilibrium. Different from the prior random-mixing analysis, here we fixed group size $n = 10$ and instead color-coded $\tau$. The results indicate that equations (\ref{wc_tau}) and (\ref{wd_tau}) yield nearly perfect matches with the simulations.

\begin{figure}[!ht]
 \centering
    \includegraphics[width=1\textwidth]{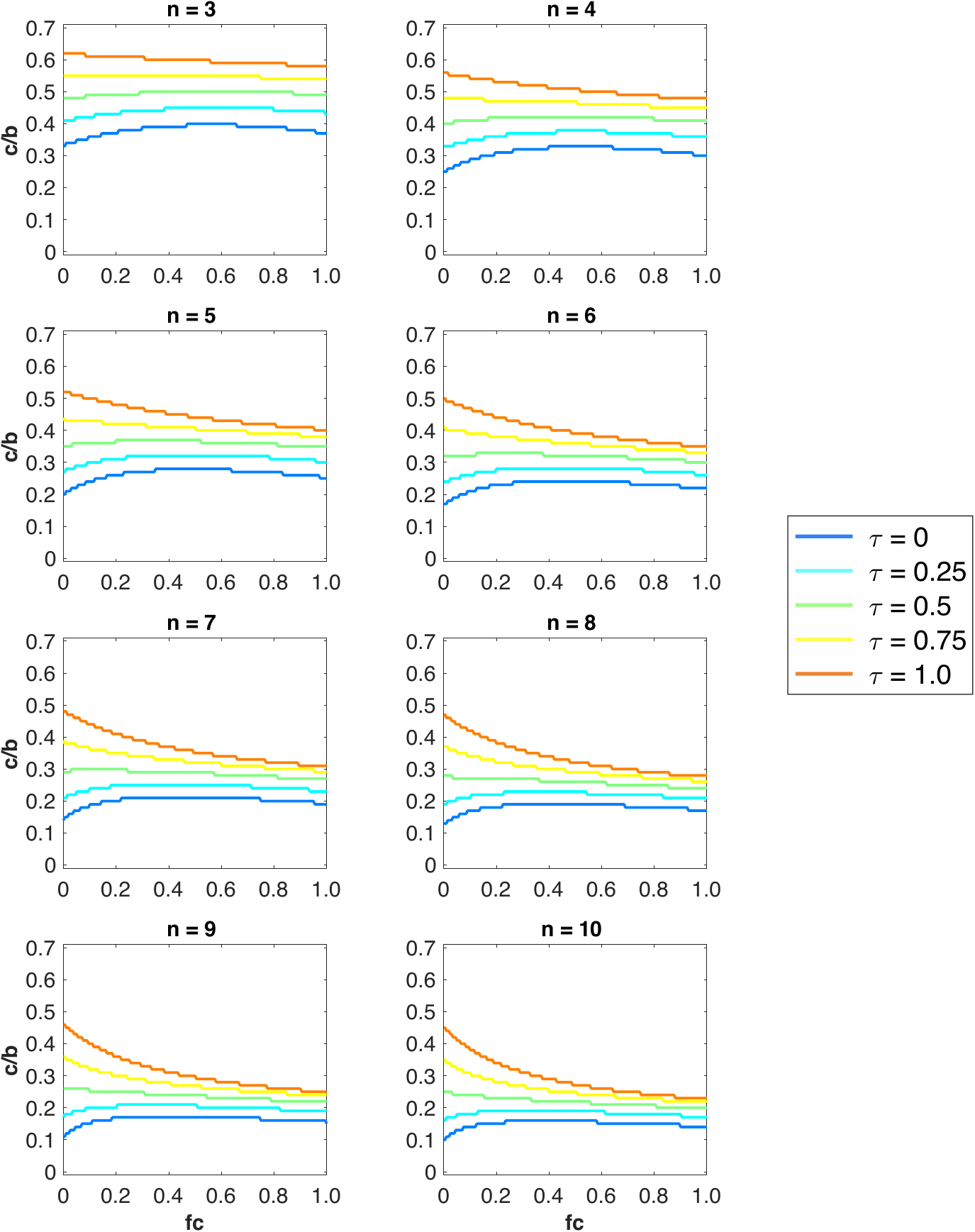}
    \caption{$W(C) = W(D)$ lines in different settings of group size $n$ and assortative level $\tau$.}
    \label{tau_equilibrium}
\end{figure}

Now, given that our analytical equations match the agent-based simulations, we used the equations to investigate the behavior of our model. The results in Figure \ref{tau_equilibrium} suggest that, across group sizes, the internal equilibrium between cooperation and defection increases as $\tau$ increases, supporting Mark's general observation that assortativity protects cooperation against defection at any given level of the cooperator-to-defector ratio. Looking closely, our model further shows that the ability of assortativity to heighten this equilibrium, though still existing, decays as $f_{c}$ increases. Superficially, Mark's work would anticipate a similar phenomenon, because when most members in a group are cooperators, there should be little need to shield them from defectors. Assortativity becomes useless in short. However, diverging from this interpretation of the findings, not only does our model indicate that assortativity's contribution to the evolutionary strength of cooperation against defection wanes at high $f_{c}$; we found it to reduce so much to the extent that the common benefit of high $f_{c}$ is compromised and even reversed. For instance, as can be seen in the figure, when a group is fully assortative, $c/b$ in fact decreases as $f_{c}$ increases--that is, the more cooperators in an assortative group relative to defectors, the \textit{less} cooperation will evolve against defection. 

Here, we believe the cause of this trade-off between $\tau$ and $f_{c}$ can be traced back to the fact that our model allows multi-leader structures, whereas Mark's model does not. As mentioned in the above discussion of random-mixing results, our multi-leader hierarchy comes with a price: It attenuates group members' willingness to contribute and makes cooperators partial defectors. As a result, when the group is large (i.e., $n$ is big) and most of its members are cooperators (i.e., $f_{c}$ is big), multiple leaders would likely emerge among the cooperators, and the group can effectively be deemed as having more but not fewer defectors than when there are fewer leaders. Non-randomly pairing these partial defectors with one another--assorting the group--subsequently contradicts the idea of protecting cooperators from defectors; cooperators are themselves, defectors, here. Indeed, this detrimental effect of high $f_{c}$ explains why there is a backfiring effect reported above that leads to the inverted U-shape equilibrium for randomly mixed structures yet no inverted U observed here. This is not about the eventual changing of the direction of the influence of $f_{c}$, either changing up or down, but about its consistent risk to mitigate cooperation when being too high, either for non-assortative or assortative groups. In summary, the present paper reports that assortativity not merely independently adds to the effect of the cooperator-to-defector ratio; the two interact. This changes the very dynamics of the evolution of cooperation in groups, where both assortativity and large numbers of cooperators can jeopardize the emergence of cooperation.

\begin{figure}[!ht]
 \centering
    \includegraphics[width=1\textwidth]{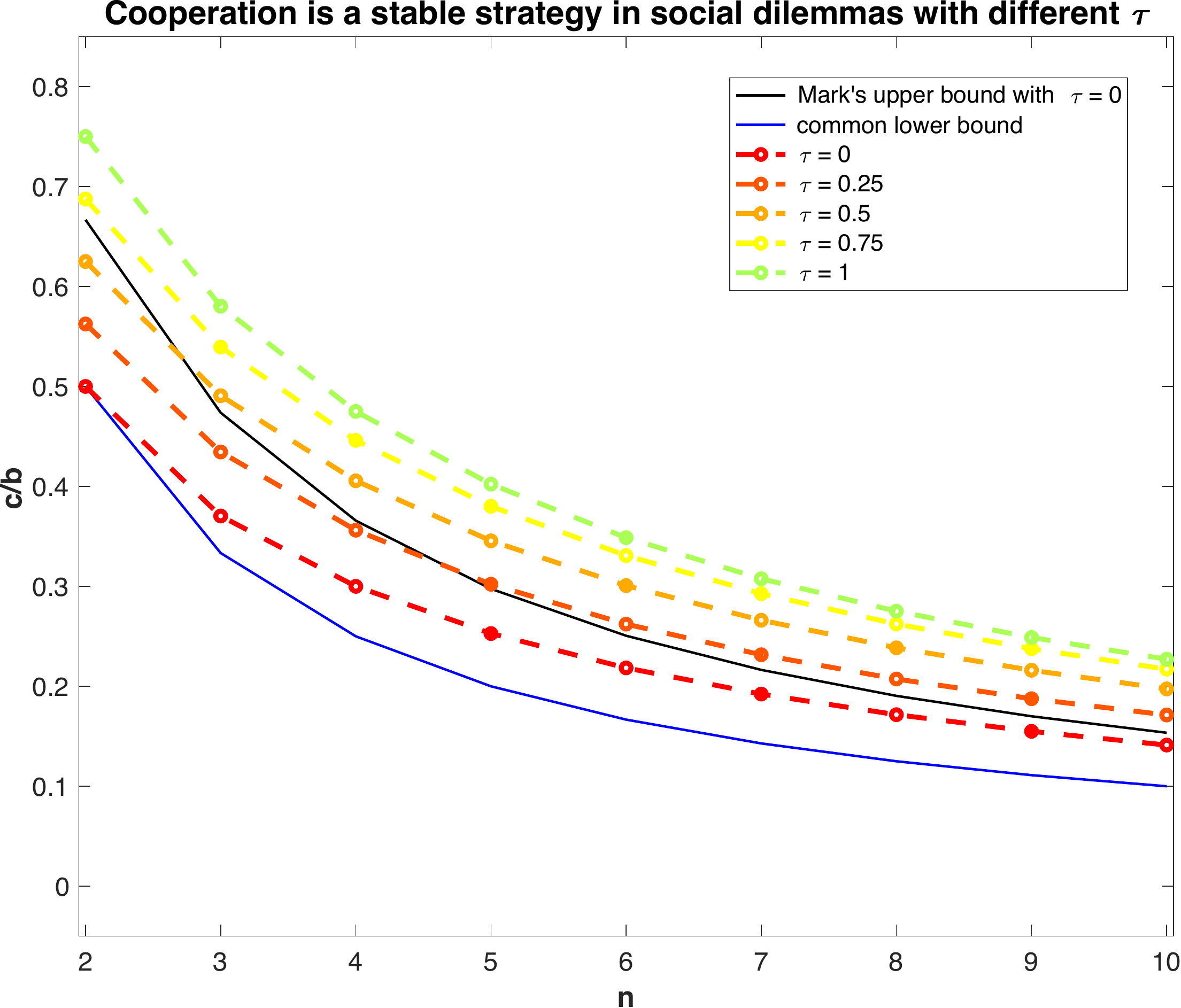}
    \caption{Cooperation is a stable strategy in social dilemmas with different assortative level $\tau$.}
    \label{tau_band}
\end{figure}

Lastly, we investigated the effects of assortative mixing on the range of cost-to-benefit ration $c/b$ for cooperation to be a stable strategy in social dilemmas. The result is shown in Figure \ref{tau_band}. Again, the lower bound of the social dilemma is the same as for randomly mixed groups, independent of $\tau$. For the upper bounds, we added the cases of different assortative levels $\tau$. Here, the upper bound is the same as the one we obtained in the random-mixing case when $\tau = 0$. When $\tau$ increases, the upper bound also increases. One can therefore see the effect of assortative mixing on promoting cooperative behavior again.

\section{General Discussion}

Using evolutionary game theory, Mark (2018) provided a pioneering evolutionary account for how group cooperation may benefit from status hierarchy. With both analytical and numerical examination, we present am extension to a single-leadership model by gauging the hierarchicalness of these multiple-leader cases and then estimating the level of cooperation mobilized by the leader-``s." We show that, when these multiple-leader scenarios are taken into account and properly handled, the emergence of cooperation against the invasion of defection in fact seems stronger than predicted by Mark's original calculation. In other words, our work not only generalizes Mark’s model over the continuous spectrum of hierarchicalness, but also conveys an optimistic message regarding how group cooperation may go hand-in-hand with status hierarchy. Indeed, to verify the analytical results of our proposed model, we developed an agent-based model, ran computer simulation using the model over tens of thousands of random replications, and demonstrate that the results of simulation well match those of our calculation. Since we have publicized the simulation code (programmed in MATLAB) in an open repository in support of full scientific transparency and sharing, the agent-based model as well as its analytical counterpart further opens the door for future follow-up attempts to replicate and to extend our current findings.

\subsection{Empirical basis of the current model}

Despite being a theoretical and computational paper, the present research is motivated and supported by empirical evidence and experiences. First, it is trivial to say that people do live in a world with multi-leader organizations; they often if not always collaborate with one another in such environments. This fact, besides the questionable calculation of Mark's paper, serves as the reason that we conducted the current investigation. And we do find cooperation possible and conditionally stable when there are more than one leader in a group in the model, as it is in the real world. Notably, our model shows that, though still possible, the more individuals competing for leadership in a group, the less stable cooperation is a behavioral strategy against defection. This is in line with the many scientific studies reporting that, say, the marginal benefit of having star investors on an investment team not only decreases when the number of such individuals increases. With the rising challenge of finding consensus among high-status group members, having talented investors on the team eventually backfires and reduces team profits \citep{groysberg2011too}. That is, the group members would in fact be better off if not working together. Anecdotally, many might also remember the surprising bumpy start of the Miami Heat of their 2010-11 NBA season when LeBron ``King" James first joined with two other all-stars on the team, Dwyane Wade and Chris Bosh. Without having played together enough, the ``Big Three" seemed to lead the team independently as three individuals as opposed to working as one unit, resulting in many lost games. Indeed, commentators regularly attribute the Heat's successes in winning the championship in the following years to the three's finally finding a way to, literally, play at different positions in one and only united system, again implying that the more ``independent" leaders, the less cooperation may achieve.

In addition to predicting what people experience in everyday life, our model relies on assumptions in line with everyday experiences too Particularly, we assume that the more group members, the less they would want to be the leader. This follows from the established bystander effect in social psychology such as that shown in Darley and Latané's \citep{darley1968bystander} research. In this classic series of experiments, Darley and Latané demonstrate that study participants were less likely to speak up and report potential emergencies--fire in the laboratory for example--when there were more participants in the lab. In other words, even if presumably everyone wanted to get out of the fire, when the group was large, few would emerge to lead the rest. By contrast, it was found easy for a person to lead themself and simply leave the room. Moreover, we build our major contribution to the literature--that cooperation is possible not only in single--but also in a multi-leader hierarchy--on the premise that the more leaders in the group, the less likely the rest would follow. The design is in accordance with the long-standing Competing Values Framework of leadership, which states that the leadership of an organization often has competing goals. The goals can be of a single leader and need to be reconciled within themself, or of different leaders and need to reconcile among themselves. Either way, if the goals cannot be integrated within or between leaders, the resources of the organization would be divided into the goals and thereof used less effective for all goals. For instance, it is reported that in a politically polarized country--the U.S. in this study--people tend to be less cooperative with those who follow a different political ideology even on non-political daily economic issues \citep{mcconnell2018economic}. As a potential result, human resources in general in the country would be arranged along the party line and, as in our model, inefficiently organized to achieve the common good for all. From here, consequently, not only does our theorizing stem from empirically bolstered bases; another way to conceptualize our work is to think of high-status agents in our model not as leaders, but as leaderships. Instead of treating them as individual persons who may or may not hold contradictory opinions, we define the agents as paradoxical opinions that the group has to juggle and allocate resourceful cooperation between, thus opening the implications of the current paper for information processing and integration within groups and even within individuals' minds.

\subsection{Implications}

One important contribution of the current paper is that we go one step further to show that cooperation can be a stable strategy even if the leadership in a group is shared by many and thus attenuated. For theory, we hold that this advance well increases the realism of our model compared to Mark's, as organizational leadership is often shared than monopolized in the wild. Even if it is held by only one person in some cases--say, of presidential power--the process of forming such clear consensus is commonly costly and, indeed, painful. Just think about how a politician funnels through the countless public debates and votes to survive their party nomination to first become the candidate, and then more debates and votes to finally win the one presidential seat. All these hurdles exist exactly because there can only be one president. Ignoring the costs resulting from this kind of challenge in the leader selection processes, especially in single-leader structures such as those in Mark's model, consequently, seriously compromises the applicability of the model. Putting the hidden costs back in sight as in our model, on the other hand, serves as a significant step closer to formal modeling the negotiation of leadership in groups and--in the present paper--its influences on cooperation among group members. Finally and importantly, we demonstrate that cooperation can emerge from multi-leader models as well, even if the leader's authority is divided among multiple individuals, just as in the real world; people regularly if not always work with one another for the common good, under the direction not merely of one, but also of many \citep{dust2016multi}. The validity of our analysis is therefore bolstered by daily observations; its future generalizability is then laid out.

As for generalizability, nevertheless, we would like to point out one implication from the present work that we believe still needs future scrutiny: that cooperation under hierarchy may be more stable than predicted by Mark's original model. On the one hand, it is encouraging to see cooperation thrive. This is what we hope for humanity and what scientifically aligns with everyday experiences--people do cooperate from time to time in a multi-leader world. On the other hand, we are uncertain about jumping to the conclusion that cooperation may emerge more easily in a universe where leadership can be divided into groups--that is, our model--than when it cannot be--Mark's model. The reason is we ourselves find that cooperation is more stable and stronger against defection when the hierarchy is clearer with fewer leaders at the top sharing the authority. This is tested analytically as well as numerically and is indeed in line with Mark's theorizing. As such, following his erroneous calculation by which all structures beginning with multiple leaders ended up having exactly one leader, all multi-leader, less-cooperation-supporting hierarchies in our model would be treated as single-leader and, therefore, fully-cooperation-supporting in Mark's model. Consequently, Mark's original prediction should have rendered cooperation more but not less stable than are our mathematical as well as agent-based prediction, which accords among themselves. Here, we are reluctant to chase the discrepancy down the rabbit hole; it is not the aim of the current work. We are however welcome future endeavors on the general issue of the effect of attenuated leadership on cooperation, hoping our model and computer code fully publicized help with the investigation.

\subsection{Limitations and future directions}

Multiple directions can be considered to extend our current work. First, both our and Mark’s models have merely considered the 2-level hierarchy (with only the top and the bottom level) but clearly, social hierarchy in real life takes various other forms. Intuitively, for instance, the most hierarchical structure is the linear system, wherein the top actor outranks the second, who in turn outranks the third, and so on and so forth until the bottom position; this system can easily have more than two levels. Further, however, prevalent linear systems are in the dominance structure of social animals \citep{shizuka2012social}, they still do not characterize all human status structures. For one, in many contests such as those of sports, while high-ranked contestants by definition win more in a tournament, they may still lose to low-ranked others from time to time. In ethology, researchers have also found cases (e.g., gorillas) where dominance relations in the group are non-transitive in the sense that actor A dominates B and B dominates C, but C somehow turns back to dominate A, thus forming a cycle structure in the group. Together, future work may continue to the literature by investigating the extent to which these diverse hierarchical structures ameliorate or deteriorate the evolution of group cooperation and, importantly, how the structures affect cooperation. 

In terms of the mechanisms through which hierarchy may benefit cooperation, we see two lines of research that may shed light on this investigation. First, besides examining the within-group competition between status cooperators and defectors, the level of analysis can be shifted up to the group level to focus on how, for instance, the cultural contexts embedding groups with different levels of emphasis on status egalitarianism influence cooperation within groups. Social norms are arguably one of the strongest propellants for the evolution of group cooperation \citep{herrmann2008antisocial}. Combining the cultural evolution models \citep{henrich2012culture}, future modeling efforts can, therefore, examine the effects of larger social norms on both status hierarchy and group cooperation. This line of investigation will extend our current work to the co-evolution of social hierarchy, group cooperation, and culture. 

Further, one may dig deeper down into individual persons' minds to study the psychology that associates being in a hierarchical situation at the moment with the decision to or not to act cooperatively. In the management and the organizational psychology literature, scholars have shown that, although at the moment subordinates may follow orders and cooperate with one another regardless, the motivating reasons that they follow orders still have a significant impact on whether the cooperation would persist in the future. For instance, if one contributes to their group, feeling autonomous in doing so, cooperation may be more likely to continue than it may when one feels forced \citep{maner2016dominance}. Even if similarly willing, those who approach the decision from an exchange-orientated perspective have been further found to be motivated to contribute more easily yet, in the meanwhile, be demotivated more easily as well and ``pull out" more quickly in future, than are those approach the decision of cooperation from a communal-orientated perspective \citep{thompson1998relationships}. 

Here, especially pertinent to our present research on leader hierarchy is the evidence that followers' psychology often reflects the psychology of their leaders. For example, the exchange orientation of subordinates can be caused by the transactional leadership of their supervisors. By contrast, subordinates' communal orientation can result from their supervisors' transformational leadership \citep{wang2005leader}. Together in the terminology of EVG, it is hence the case that the psychology underlying cooperation behavior influences the strength of this behavioral strategy for shielding itself from the invasion of defection and, therefore, its stability over the long run. When combined with the model proposed in the current paper, this direction for future work may then broaden the scope of leader ``psychology" types in the formal modeling of cooperation in hierarchical organizations.

Finally, corresponding to the great theoretical variety of hierarchy is the large literature on the methodology of measuring hierarchicalness. While one can always try a different measure to make their model perform better, not all of the alternatives would work equally well. For instance, as does Mark’s model, our model does not take into account information about individuals differentials in the social hierarchy. This makes the first kind of hierarchicalness measure discussed in section two (i.e., social hierarchy and cooperation) becomes inapplicable. Even if other methods are logically applicable--take the measure by \citep{krackhardt1994graph} for example--while the measure shares a similar idea with the method developed by \citep{mones2012hierarchy} and adopted by the present paper, only the latter but not the former distinguishes the hierarchicalness of a structure--say, one that has one leader and $n-1$ subordinates--from the hierarchicalness of the structure's reverse form--one that has $n-1$ leaders and one subordinate. In other words, Kackhardt’s method would treat the two structures as equally hierarchical, which is against Mark’s theorizing and clearly counter-intuitive, since the single-leader structure is commonly deemed more hierarchical than the single-subordinate one. In contrast, the measure by \citep{mones2012hierarchy} works better in pinpointing this asymmetry between the two example structures, and thus is more suitable for the purpose of our work. Overall, then, future investigations would want to be mindful of both the theory and the methodology, as well as of their fit with each other. Just because our choice of the hierarchicalness measure functions well in our research does not mean it will do the same elsewhere. We encourage researchers to explore their options in the future.

\clearpage

\section{Appendix}

\subsection{Revisit of the average payoff of status cooperators and defectors}

In this subsection, we revisit Mark's original model of the average payoff of status cooperators and defectors. Here in the signaling process, if the group reaches a status hierarchy, only one individual could be on the top level. If more than one status cooperators go to the top level, this yields a failed attempt and the group starts its signaling process again. Note that in this setting, it is possible that during the first attempt there are more than one status cooperators at the top level. But since this is a failed attempt, later there could be no status hierarchy, i.e., no individuals at the top level.

Let $f_{c}$ be the ratio of status cooperators in a group, the formula (4a) and (4b) in Mark's paper become:

\begin{equation}
\begin{aligned}
\label{wc_new}
W(C) &= \sum^{n-1}_{i = 0} {n-1 \choose i} (f_{c})^{i} (1-f_{c})^{n-1-i} \\
& \cdot \bigg[ \frac{{i+1 \choose 1} (\frac{1}{n}) (\frac{n-1}{n})^{i}}{{i+1 \choose 1} (\frac{1}{n}) (\frac{n-1}{n})^{i}+(\frac{n-1}{n})^{i+1}} \cdot \bigg(\frac{(i+1)b}{n}\bigg)+ \frac{(\frac{n-1}{n})^{i+1}}{{i+1 \choose 1} (\frac{1}{n}) (\frac{n-1}{n})^{i}+(\frac{n-1}{n})^{i+1}}\cdot c\bigg],
\end{aligned}
\end{equation}
and 
\begin{align}
\label{wd_new}
W(D) &= c + \sum^{n-1}_{i = 0} {n-1 \choose i} (f_{c})^{i} (1-f_{c})^{n-1-i} \cdot \bigg[ \frac{{i \choose 1} (\frac{1}{n}) (\frac{n-1}{n})^{i-1}}{{i \choose 1} (\frac{1}{n}) (\frac{n-1}{n})^{i-1}+(\frac{n-1}{n})^{i}} \cdot \bigg(\frac{ib}{n}\bigg)\bigg].
\end{align}

\begin{figure}[!ht]
 \centering
    \includegraphics[width=1\textwidth]{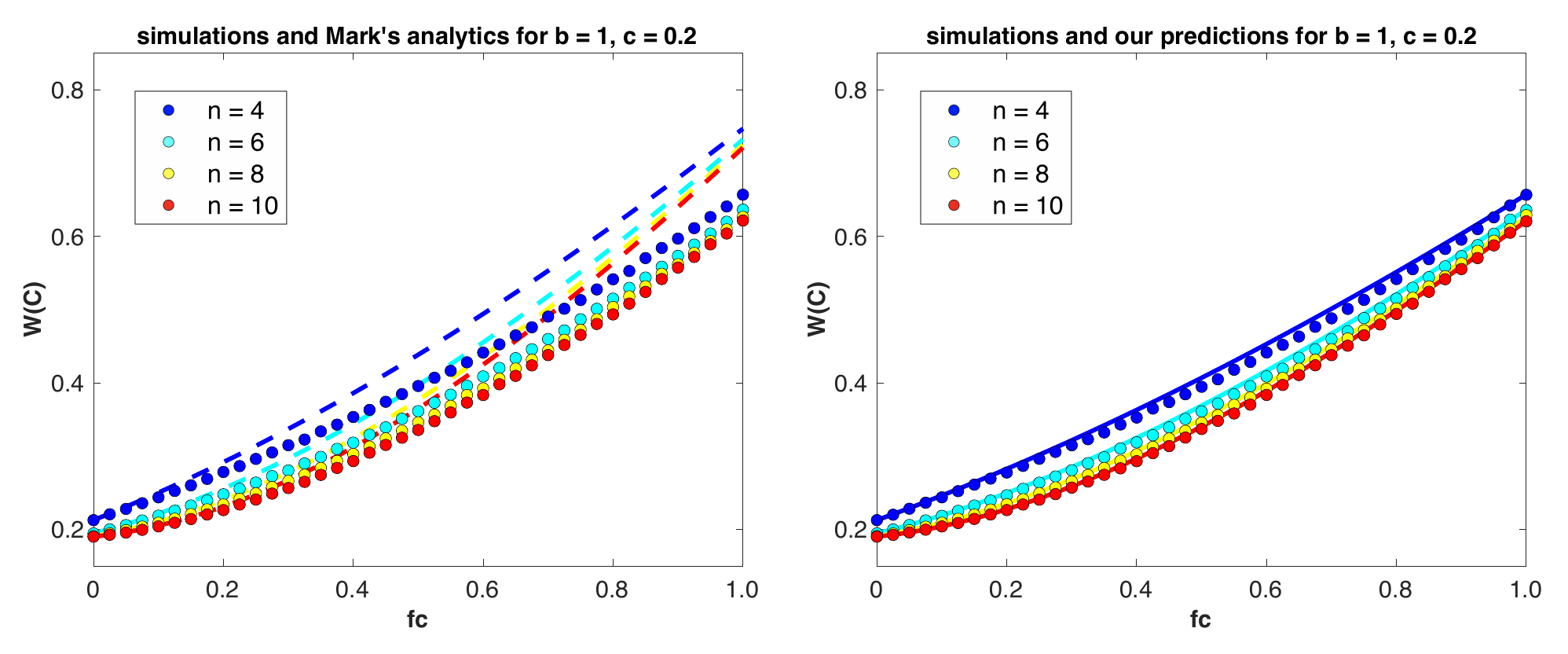}
    \caption{Comparison of W(C) by simulating form Mark's original model using formula (4a) (left, dashed lines) and our formula (right, solid lines).}
    \label{fig_wc}
\end{figure}

\begin{figure}[!ht]
 \centering
    \includegraphics[width=1\textwidth]{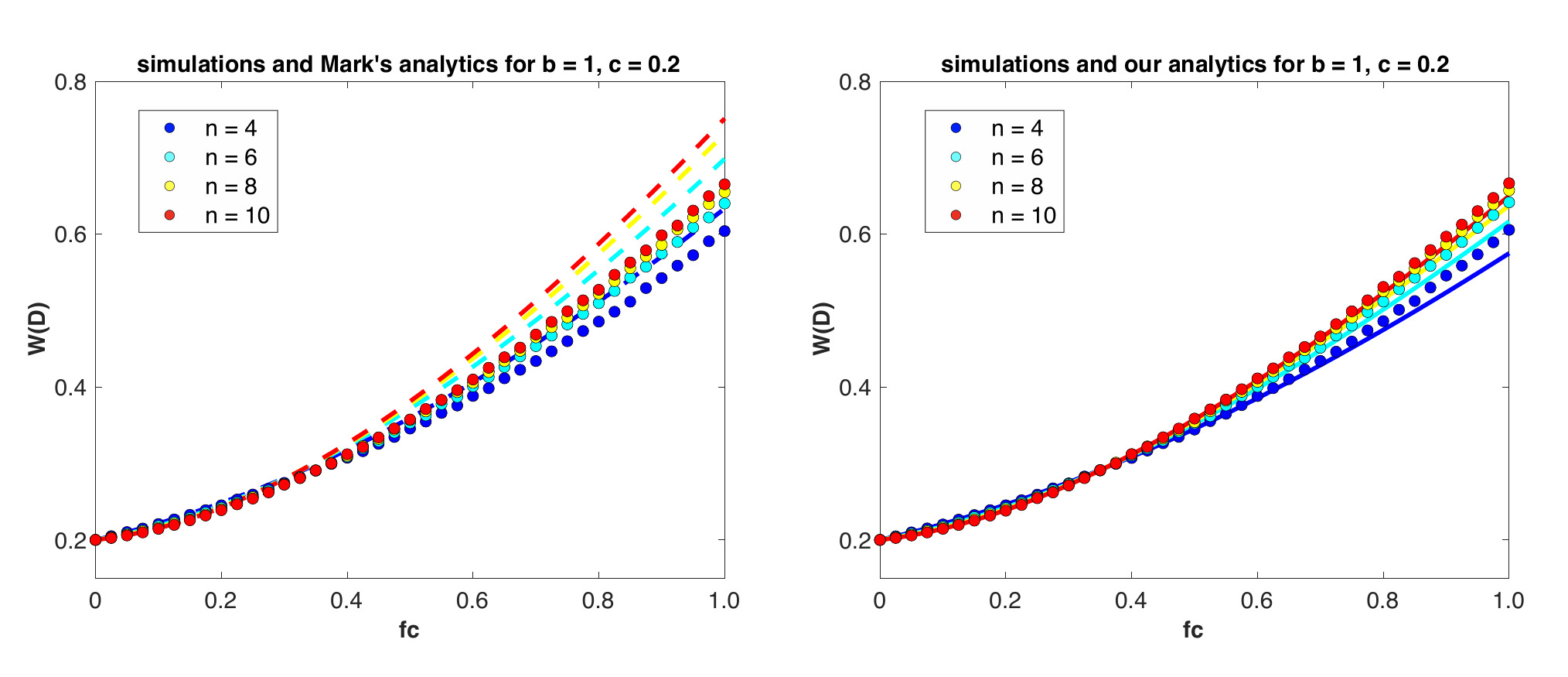}
    \caption{Comparison of W(D) by simulating form Mark's original model using formula (4b) (left, dashed lines) and our formula (right, solid lines).}
        \label{fig_wd}
\end{figure}

Figure \ref{fig_wc} and \ref{fig_wd} are the comparison of simulations, Mark's formula, and our formula. We build and agent-based model for the phase of preplay signaling and payoff determination. In order to compute expected values of $W(C)$ and $W(D)$, the parameters we choose here are $b = 1$ and $c = 0.2$. Each dot is an average of 100,000 simulations. We can see using Mark's formula (4a) and (4b), the prediction gets worse when $f_{c}$ is large. The discrepancy in Mark's formula and simulations is not getting smaller as we increase the group size $n$. Note that there are still discrepancies for our predictions, but as $n$ increases, the discrepancies become smaller asymptotically.

\subsection{Simulations and analytic codes}
All the codes are available on the author's GitHub account: \\
\url{https://github.com/waynelee1217/status\_cooperation}

\clearpage

\bibliographystyle{apalike}
\bibliography{references}

\end{document}